\documentstyle[preprint,aps,epsf,floats]{revtex}
\renewcommand{\baselinestretch}{1.5}
\begin{document}
\begin{titlepage} 
\renewcommand{\baselinestretch}{1}
\small\normalsize
\begin{flushright}
hep-th/0006099\\
MZ-TH/00-25     \\
\end{flushright}

%\vspace{1cm}

\begin{center}   

{\LARGE \sc Rotation Symmetry Breaking Condensate\\
 [5mm]
 in a Scalar Theory}

\vspace{1cm}
{\large O. Lauscher\\
\vspace{0,2cm}
M.\ Reuter} \\
\vspace{0.5cm}
\noindent
{\it Institut f\"ur Physik, Universit\"at Mainz\\
Staudingerweg 7, 55099 Mainz, Germany}\\

\vspace{1cm}
{\large
C. Wetterich}\\
\vspace{0.5cm}
\noindent
{\it Institut f\"ur Theoretische Physik, Universit\"at Heidelberg\\
Philosophenweg 16, 69120 Heidelberg, Germany}
\end{center}   
 
\vspace*{0.5cm}
\begin{abstract}
Motivated by an analogy  with the conformal factor problem in gravitational
theories of the $R+R^2$-type we investigate a $d$-dimensional Euclidean field
theory containing a complex scalar field with a quartic self interaction and
with a nonstandard inverse propagator of the form $-p^2+p^4$. Nonconstant 
spin-wave configurations minimize the classical action and spontaneously break
the rotation symmetry to a lower-dimensional one. In classical statistical 
physics this corresponds to a spontaneous formation of layers. Within the 
effective average action approach we determine
the re\-nor\-ma\-li\-za\-tion group flow of the dressed inverse propagator and
of a family of generalized effective potentials for nonzero-momentum modes. 
Already in the leading order of the semiclassical expansion we find strong 
``instability induced'' renormalization effects which are due to the fact that
the naive vacuum (vanishing field) is unstable towards the condensation of 
modes with a nonzero momentum. We argue that the (quantum) ground state of our
scalar model indeed leads to spontaneous breaking of rotation symmetry.
 \end{abstract}
\end{titlepage}
%
%\newpage
\section{Introduction}
\label{1}
It is a common feature of several Euclidean field theories of physical interest
that spatially inhomogeneous, i.e. nonconstant field configurations have a
lower value of the action functional than homogeneous ones. This means that,
at least semiclassically, the inhomogeneous configurations are likely to
dominate the functional integral and thus to determine the quantum vacuum
state $\left|0\right>$. In this case we expect that the essential features of 
the true vacuum state can be understood by an expansion in the quantum 
fluctuations about a set of 
configurations with a position-dependent, non-translational invariant value of
the field variable. In the quantum vacuum this ``condensation'' of spatially 
inhomogeneous modes contributes to certain expectation values $\left<0\right|
{\cal O}\left|0\right>\neq 0$ where ${\cal O}$ is a scalar 
operator constructed from the derivatives of the fundamental fields in such a
way that it is sensitive to the nonvanishing kinetic energy of the contributing
configurations. (For instance, in a scalar model, ${\cal O}=\partial_\mu\phi\,
\partial^\mu\phi$.) We shall generically refer to such contributions as
``kinetic condensates''. They are to be distinguished from the familiar
translational invariant ``potential condensates'' characterizing the 
conventional Higgs mechanism which is triggered by a nonzero but constant 
scalar field expectation value.

In this context one should distinguish two different physical situations. If 
the degenerate minimum of the effective action corresponds to configurations
which are not invariant under some global symmetry like translations or 
rotations such a symmetry is spontaneously broken and only the remaining 
unbroken symmetry can be used for a classification of the spectrum of 
excitations. This spectrum will typically contain massless 
``Goldstone-excitations''. In this case one typically has an order parameter
whose expectation value breaks the symmetry, in addition to the invariant 
kinetic operators mentioned above. In the second case the dominant 
configurations break a local symmetry. Then it is well known that there is no
true spontaneous symmetry breaking and no nonzero expectation value of a
noninvariant order operator exists. Also the ``Goldstone-excitations'' are
absent from the physical spectrum. Nevertheless, we have learned from the
Higgs-mechanism in the electroweak standard model that a language in terms of
``spontaneous symmetry breaking'' can be very useful. This spontaneous
symmetry breaking manifests itself in nonperturbative contributions to 
invariant operators for ``kinetic condensates''.

A typical example of a kinetic condensate is the gluon condensate in QCD.
While the classical Yang-Mills action $1/4\int d^4x F_{\mu\nu}^a F_{\mu\nu}^a$
is minimized by $F_{\mu\nu}=0$, already the one-loop effective action assumes
its minimum for $F_{\mu\nu}\neq 0$. The Savvidy vacuum \cite{sav} tries to 
model the true ground state as a covariantly constant color magnetic field. 
The effective action of this state is indeed lower than that for 
$F_{\mu\nu}=0$. It is known, however, that the Savvidy vacuum is unstable in 
the infrared (IR), and it has been speculated that the dominant configurations
may be spatially inhomogeneous (perhaps domain-like) in order to provide an IR
cutoff at a scale set by $\Lambda^{-1}_{\rm QCD}$. Those nontrivial properties
of the QCD vacuum are parametrized\footnote{Invariant operators like $\left<0
\right|F_{\mu\nu}^a F_{\mu\nu}^a\left|0\right>$ receive also perturbative 
contributions which are not related to ``kinetic condensates''. We do not 
discuss here the difficult problem how they can be separated from ``kinetic
condensates''.} by $\left<0\right|F_{\mu\nu}^a F_{\mu\nu}^a\left|0\right>$ and
similar condensates of more complicated gauge- and Lorentz-invariant operators.

Another important example is Euclidean gravity based upon the Einstein-Hilbert
action \cite{haw}
\begin{eqnarray}
\label{m0}
S_{\rm EH}\left[g_{\mu\nu}\right]=-\frac{1}{16\pi G}\int d^4x\,\sqrt{g}\,R
\end{eqnarray}
which is not positive definite. In fact, decomposing the metric as $g_{\mu
\nu}=\exp(2\phi)\bar{g}_{\mu\nu}$ where $\bar{g}_{\mu\nu}$ is a fixed reference
metric, we obtain
\begin{eqnarray}
\label{m1}
S_{\rm EH}\left[g_{\mu\nu}\right]=-\frac{1}{16\pi G}\int d^4x\,\sqrt{\bar{g}}
\,e^{2\phi}\,\left[\bar{R}+6\,\bar{g}^{\mu\nu}\,\bar{D}_\mu\phi\,\bar{D}_\nu
\phi\right]\,.
\end{eqnarray}
This shows that $S_{\rm EH}$ can become arbitrarily negative if the conformal
factor $\phi(x)$ varies rapidly enough so that $(\bar{D}_\mu\phi)^2$ is large.
Leaving aside for a moment the well known pro\-blems in setting up a consistent
theory of quantum gravity, it is tempting to speculate that the theory cures 
this instability caused by the unboundedness of $S_{\rm EH}$ in a dynamical
way. Nonconstant $\phi$-modes could condense in such a way that the resulting
quantum vacuum state is stable and constitutes the absolute minimum of some
yet unknown {\it effective} action functional. The expectation value of the
metric in this state should be close to the metric for flat space (which is 
not the minimum of $S_{\rm EH}$!). Also, operators like $(\bar{D}_\mu\phi)^2$,
appropriately covariantized, should acquire nonzero expectation values. This
would indicate that the true vacuum arises from a ``dynamical stabilization''
of the bare theory due to the condensation of nonconstant $\phi$-modes.

A further model in which the existence of a variant of the ``kinetic 
condensate'' has been speculated about is Liouville field theory \cite{liou},
\cite{liou2}.
The expectation value of its operatorial equation of motion reads
\begin{eqnarray}
\label{i0}
\left<0\left|\Box\phi\right|0\right>+\frac{m^2}{\beta}\,\left<0\left|
e^{\beta\phi}\right|0\right>=0\,.
\end{eqnarray} 
Provided it is possible to make the operator $\exp(\beta\phi)$ well defined
and that the regularized operator is still positive definite, 
$\rm eq.\,(\ref{i0})$
implies that $\left<0\left|\Box\phi\right|0\right>$ is nonzero and, as a
consequence, that the vacuum $\left|0\right>$ is not translational invariant.

In the examples mentioned above the determination of the vacuum state is a 
formidable task which has not been mastered yet. In the present paper we shall
therefore study the formation of a ``kinetic condensate''
within the framework of a scalar toy model. On the one hand, this model is
simple enough to be treated analytically, on the other hand it is found to
have the feature of a ``dynamical stabilization'' which we hope to
occur in QCD and in quantum gravity. 

\begin{figure}[ht]
\hbox to\hsize{\hss
       %\vspace{0.5cm}
        \epsfxsize=8cm
        \epsfysize=6cm
        \centerline{\epsffile{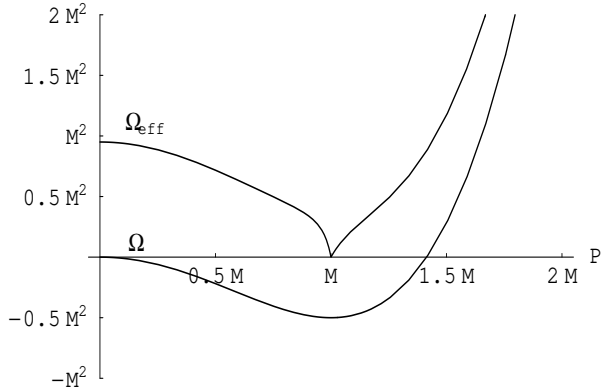}}\hss}
\caption{%
Classical inverse propagator $\Omega$ and quantum inverse 
propagator $\Omega_{\rm eff}$ (for zero background field) as functions of
${\rm p}\equiv|p|$.}
\label{f13}
\end{figure}

The model is formulated in $d$ Euclidean dimensions. It contains a massless 
complex scalar field $\chi$ with a conventional $\lambda|\chi|^4$-self 
interaction but with a higher-derivative kinetic term: 
\begin{eqnarray}
\label{a1}
S\left[\chi\right]
=\int d^dx\,\left\{\chi^*\,\Omega(-\Box)\,\chi
+\frac{\lambda}{2}\,|\chi|^4\right\}
\end{eqnarray}
The kinetic operator $\Omega$ is taken to be
\begin{eqnarray}
\label{a3}
\Omega(-\Box)=\Box+\frac{\Box^2}{2M^2}\;,\;\;\Box=\partial_\mu\partial_\mu\;,
\end{eqnarray}
so that in momentum space
\begin{eqnarray}
\label{a4}
\Omega(p^2)=-p^2+\frac{(p^2)^2}{2M^2}
\end{eqnarray}
where $M$ is a constant with the dimension of a mass. On a Euclidean
spacetime where $p^2\equiv p_\mu p_\mu\ge 0$, the kinetic operator is positive
for $p^2>2M^2$ but negative for momenta between $p^2=0$ and $p^2=2M^2$. It
has a minimum at $p^2=M^2$ where it assumes the value $-M^2/2$, see FIG. 
\ref{f13}. The action (\ref{a1}) has a global {\sf U(1)}-invariance under 
phase rotations $\chi\rightarrow\chi\,\exp(i\varphi)$ with a constant 
$\varphi$, and it is invariant under the Euclidean Poincar\'{e} group 
{\sf ISO($d$)} of rigid spacetime translations and rotations.

We will see that in this model the rotation symmetry is spontaneously broken
whereas a modified translation symmetry is preserved. As a classical 
statistical system in $d=3$ dimensions or the (zero temperature) ground state 
of a quantum statistical system ($d=4$) this models the spontaneous formation 
of two-dimensional
layers which break the rotation symmetry. An effective translation symmetry
rotates the phase factor of the complex field $\phi$ by $2\pi$ as one
translates from one layer to the next. In two dimensions it corresponds to 
the spontaneous generation of line-like structures. We emphasize that the 
microscopic action has rotation and translation symmetry, in contrast to 
lattice models. Our model therefore describes situations where already a tiny 
perturbation of these symmetries results in nontrivial geometric structures.

The model
shares some essential features with the conformal sector of a gravitational
model of the type $S[g_{\mu\nu}]=\int d^dx\,\sqrt{g}\{\alpha\,R+\beta\,R^2\}$,
for instance. (Here ``$R^2$'' stands for any invariant quadratic in the
Riemann tensor.) The Euclidean classical action of this model is bounded below,
in contrast to the Einstein-Hilbert action. It is therefore a good starting 
point for the definition of a Euclidean functional integral if the problems of
UV regularization can be mastered. The Einstein-Hilbert term $\sqrt{g}R$ leads
to a negative
contribution to the kinetic term of the conformal factor of the metric, which
dominates at small momenta, while the $R^2$-term gives rise to a positive
contribution dominating at large momenta. The instability at small and the
stability at large momenta is modelled by the ansatz (\ref{a4}) with $M$
playing the role of the Planck mass.

The ``wrong sign'' $p^2$-term in (\ref{a4}) induces an instability of the
naive vacuum with $\chi=0$ towards the formation of a spatially inhomogeneous
ground state because the system tends to lower its Euclidean action by making
the kinetic action $\int_p\Omega(p^2)|\tilde{\chi}(p)|^2$ as negative as
possible. Thus we expect that the vacuum of this model is dominated by
nonconstant field configurations whose typical momenta are of the order of the
scale $M$. We shall see that this is actually the case. 

The true vacuum configuration $\left<\chi(x)\right>$ of any theory can be
found from its (standard) effective action $\Gamma[\phi]$ by solving the
``dressed'' field equation $\delta\Gamma/\delta\phi=0$. In this paper we 
consider $\Gamma$ as the zero-cutoff limit of the effective average action 
$\Gamma_k[\phi]$, a type of coarse grained free energy with a variable
infrared (IR) cutoff at the mass scale $k$ \cite{avact}. It satisfies an exact
renormalization group equation, and it interpolates between the classical
action $S=\Gamma_{k\rightarrow\infty}$ and the standard effective action
$\Gamma=\Gamma_{k\rightarrow 0}$. For our model we shall determine the 
renormalization group trajectory $k\rightarrow\Gamma_k$ in the leading order
of the semiclassical expansion which, as we shall argue, provides a 
qualitatively correct picture already. 

The functional $\Gamma_k[\phi]$ has the same {\sf U(1)} and {\sf ISO($d$)} 
symmetry as the classical action. In particular, the bilinear term of 
$\Gamma\equiv\Gamma_0$ has
the same structure as the one in $S$ but $\Omega(p^2)$ is replaced by the
dressed  inverse propagator $\Omega_{\rm eff}(p^2)$. In FIG. \ref{f13} we have
plotted our result for $\Omega_{\rm eff}$. Quite remarkably, due to the 
renormalization effects, the kinetic term has become positive semidefinite 
even for a vanishing ``background field'' $\phi$.
For all modes with $p^2\neq M^2$ it ``costs'' energy (action) to excite 
them. By including renormalization effects these modes have been stabilized in
a dynamical way. On the other hand, modes with $p^2=M^2$ can be excited 
``for free''. This indicates that those modes might be unstable towards the
formation of a condensate.
\renewcommand{\baselinestretch}{1}
\small\normalsize
\renewcommand{\baselinestretch}{1.5}
\small\normalsize

We shall analyze this phenomenon in terms of a family of generalized momentum
dependent effective potentials $U_k(A;p^2)$ which are obtained by evaluating
$\Gamma_k[\phi]$ for plane-wave arguments $\phi=A\,\exp(ip_\mu x_\mu+i\beta)$.
It turns out that for $k\rightarrow 0$ and $p^2\neq M^2$ all $U_k$'s have their
minimum at $A=0$ so that the corresponding modes do not acquire a vacuum
expectation value. For $p^2=M^2$ the situation is different: in the limit 
$k\rightarrow 0$ the potential $U_k(A;M^2)$ develops a flat bottom which 
signals a nonzero expectation value, i.e. a condensation of the plane-wave
modes with momenta $p_\mu=M n_\mu$. Here $n_\mu$ is an arbitrary unit vector.
Indeed, for small but nonzero $k$ the absolute minimum of $\Gamma_k$ occurs
(within our approximation) for $A\neq0$ in the mode $p^2=M^2$.

In leading order we find the following expectation value of the fundamental
field:
\begin{eqnarray}
\label{w0}
\left<\chi(x)\right>=\frac{M}{\sqrt{2\lambda}}\,\exp\left(iM n_\mu x_\mu+i
\alpha\right)
\end{eqnarray}
It is characterized by the phase $\alpha$ and the vector
$n_\mu$. This means that the above expectation value leads to a spontaneous
breaking of both the {\sf U(1)} and the {\sf ISO($d$)} symmetry.
For the translations in the directions $n_\mu$ only a remaining symmetry 
corresponding to combined transformations of the form
\begin{eqnarray}
\label{w4}
x'_\mu&=&x_\mu+\frac{\xi}{M}\,n_\mu\nonumber\\
\alpha'&=&\alpha-\xi
\end{eqnarray}
is left unbroken, where $\xi$ represents a real, constant parameter. For such
combined transformations the symmetry breaking from the spacetime 
translations is compensated by an appropriate phase 
rotation so that $M n_\mu x_\mu+\alpha=M n_\mu x_\mu'+\alpha'$.
The case $\xi=2\pi m$, $m$ integer, is special in the sense that no 
compensating phase rotation is needed to achieve invariance. Therefore we 
obtain a symmetry with respect to discrete spacetime translations given by
\begin{eqnarray}
\label{w5}
x'_\mu&=&x_\mu+\frac{2\pi m}{M}\,n_\mu\,.
\end{eqnarray}
As was already mentioned above the spontaneous breaking of the
{\sf ISO($d$)} symmetry to this discrete symmetry is analogous to
the spontaneous formation of layers in statistical systems where the 
transformation (\ref{w5}) describes a translation from one layer to another.  
We emphasize that the layer structure involves the internal degrees of freedom
whereas {\sf U(1)} invariant operators have translationally invariant 
expectation values due to the combined effective translation symmetry.

By analyzing the spectrum of small fluctuations about the vacuum configuration
(\ref{w0}) we find that all those fluctuations are stable. This dynamical 
stabilization of an apparent tree-level instability is formally
analogous to what happens in the familiar situation of an ordinary kinetic
term ``$+p^2$'' along with a symmetry breaking potential $V=-\mu^2|\chi|^2+(
\lambda/2)|\chi|^4$. In this case the naive vacuum $\chi=0$ is unstable
because of the negative mass term, and shifting the field by a constant is
sufficient to reach the true vacuum. In our model the analogous ``shift to
the new vacuum'' is more involved since the field variable is shifted by an
explicitly $x_\mu$-dependent field. 

The various candidates for the ``vacuum field configuration'' can be 
distinguished by their contribution to the expectation value of $\partial_\mu
\chi^*\partial_\mu\chi$. Clearly $\left<\partial_\mu\chi^*\partial_\mu\chi
\right>=0$ for the perturbative vacuum, while we find for the ground state
(\ref{w0})
\begin{eqnarray}
\label{i2}
\left<\partial_\mu\chi^*\partial_\mu\chi\right>=\frac{M^4}{2\lambda}\,.
\end{eqnarray}
This is a translational invariant ``kinetic condensate'' quite analogous 
to the gluon condensate in QCD. The condensate (\ref{i2}) is nonanalytic
in the coupling $\lambda$, i.e. it could not be seen in any finite order of
perturbation theory.

The remaining sections of this paper are organized as follows. In section 
\ref{2} we discuss the classical vacua, i.e. the degenerate absolute minimum
of the functional $S[\chi]$, and we determine the spectrum of small 
fluctuations about those field configurations. In sections \ref{3} and 
\ref{10} we review some aspects of the average action approach, and we discuss
the phenomenon of ``classical renormalization''.
It is well known that in theories where standard perturbation theory about the
trivial vacuum $\chi=0$ is applicable the lowest order (i.e., classical) 
approximation of the loop expansion yields $\Gamma=S$ and, more
generally, $\Gamma_k=S$. We shall see that for unstable theories such as the 
one investigated here $\Gamma$ does {\it not} equal $S$ plus terms 
induced by loops. There are nontrivial renormalization effects even at the 
classical level. The reason is that in this case the loop expansion must be
performed about the true minimum of the classical action rather than the false
vacuum configuration $\chi=0$. Within the loop expansion this type of 
classical renormalization \cite{RW90} is 
refered to as ``instability induced'' \cite{ABP99} as opposed to the familiar
``fluctuation induced''  renormalization coming from the loops.\footnote{From
the point of view of the exact renormalization group equation this distinction
disappears \cite{TW92}.} In 
order to understand the full quantum theory at a qualitative level it is often
sufficient to take the classical renormalization of the effective action into
account. It encodes the physics related to the shift to the true vacuum.

The average action $\Gamma_k$ is defined in terms of a functional integral 
which contains a modified classical action $S_k^J$ containing the IR-cutoff
and source terms. The semiclassical expansion will be applied to this integral.
Therefore we need to know the absolute minimum $\chi_{\rm min}(x;J,k)$ of this
functional. In section \ref{4} we derive a sufficient condition for a 
configuration to be the global minimum of $S_k^J$. This is used in section 
\ref{5} in order to establish that for vanishing sources the minimum is
constituted by spin waves of the type (\ref{w0}). For $J\neq 0$ the situation
is more complicated and the corresponding discussion can be found in
appendix \ref{Acomplete}. 

In section \ref{6} we compute the bilinear term of $\Gamma_k$ and derive the
renormalization group flow of the dressed inverse propagator $\Sigma_k(p^2)$ 
for vanishing background field. In particular, 
at the end point $k=0$, we obtain the full quantum inverse propagator 
$\Omega_{\rm eff}(p^2)$. In section \ref{7} we identify regions in field space
with no classical renormalization, i.e. configurations $\phi$ for which 
$\Gamma_k[\phi]=S[\phi]$ is indeed true in lowest order. Then, in section 
\ref{8}, we combine
all pieces of information available in order to draw a global picture of the 
effective average action $\Gamma_k$. We introduce the momentum dependent
potentials $U_k(A;p^2)$ and use them to discuss the structure of the true
vacuum. Some calculational details are given in the appendices \ref{D}, 
\ref{C} and \ref{E}. The conclusions are given in section \ref{9}.

\section{Spin waves and their fluctuation spectrum}
\label{2}
For every Euclidean field theory the field configuration $\chi_{\rm min}$
with the smallest possible value of the action $S$ is of special importance.
For a massless $|\chi|^4$-theory with an ordinary kinetic term $\Omega=+p^2$
this configuration is $\chi_{\rm min}(x)=0$ which puts to zero the potential
and the kinetic energy separately. In our case the situation is less trivial
because $\Omega(p^2)$ can assume negative values and hence it is able to
compensate for positive contributions coming from the potential. In fact, we
shall prove later on in a more general context that the absolute minimum of 
the action (\ref{a1}) is achieved for the ``spin'' wave configurations
\begin{eqnarray}
\label{b6}
\chi_{\rm min}(x)=\frac{M}{\sqrt{2\lambda}}\,\exp\left(i M n_\mu x_\mu+i\alpha
\right)\,.
\end{eqnarray}
Here $n_\mu$ is an arbitrary unit vector (a point on $S^{d-1}$) and $\alpha$
is a free phase. Thus the global minimum is degenerate and the corresponding
``vacuum manifold'' is $S^1\times S^{d-1}$. Different points of this manifold
correspond to different classical ground states of the theory. If we pick one
of those ground states, characterized by a fixed pair $(n,\alpha)$, this
amounts to a spontaneous breaking of the {\sf ISO($d$)} symmetry of spacetime
rotations and translations as well as of the {\sf U(1)} group of global phase
transformations. 

The classical $(n,\alpha)$-vacua are characterized by a nonzero expectation
value of the operator $\partial_\mu\chi^*\,\partial_\mu\chi$, for instance.
The spin wave solution (\ref{b6}) yields 
\begin{eqnarray}
\label{d0}
\left<\partial_\mu\chi^*\,
\partial_\mu\chi\right>\equiv\partial_\mu\chi_{\rm min}^*\,\partial_\mu
\chi_{\rm min}=\frac{M^4}{2\lambda}\,.
\end{eqnarray}

\begin{figure}[ht]
\begin{minipage}{6cm}
        \epsfxsize=6cm
        \epsfysize=5cm
        \centerline{\epsffile{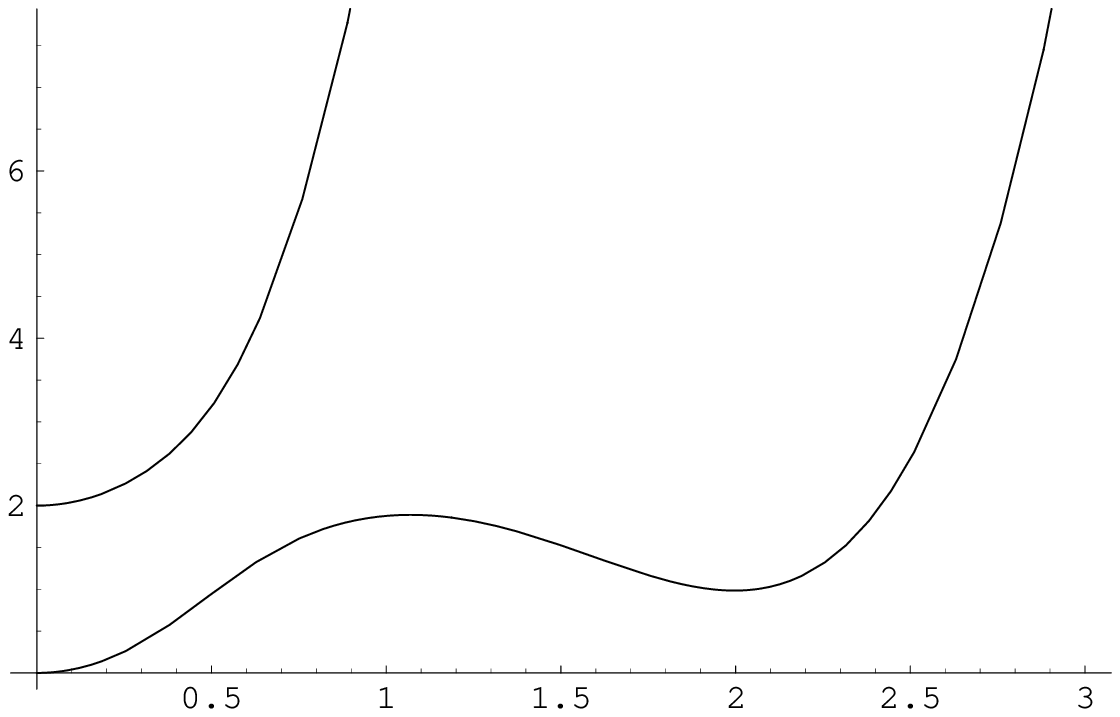}}
\centerline{(a)}
\end{minipage}
\hfill
\begin{minipage}{6cm}
        \epsfxsize=6cm
        \epsfysize=5cm
        \centerline{\epsffile{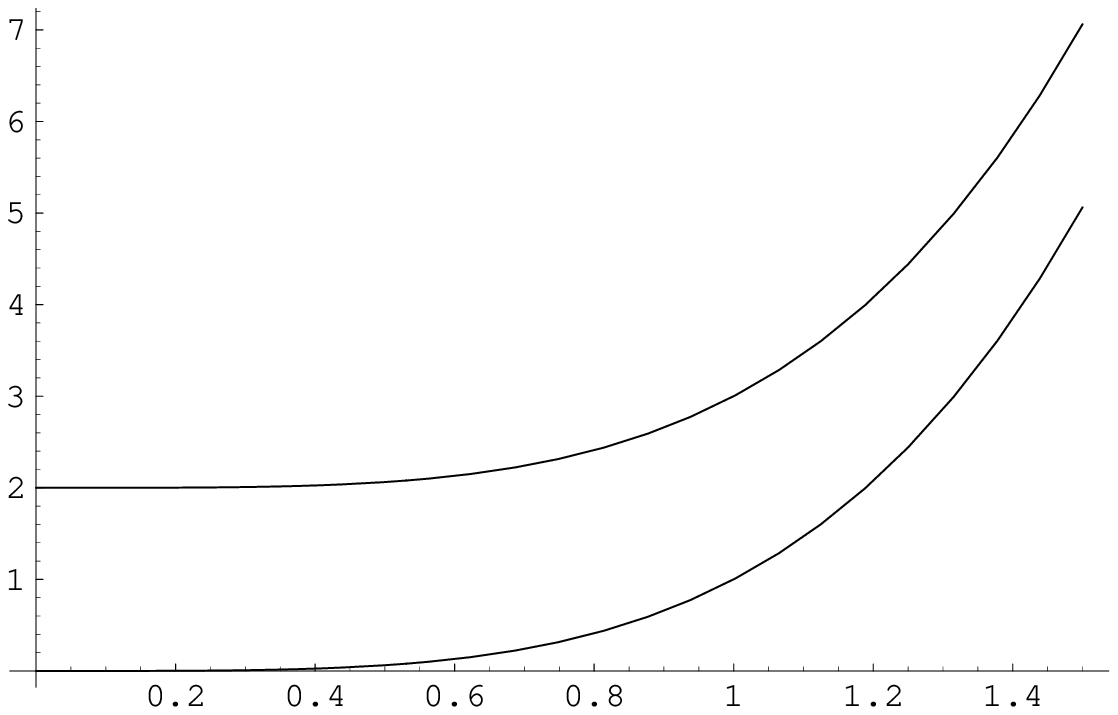}}
\centerline{(b)}
\end{minipage}
\vspace{0.3cm}
\caption{%
${\cal K}_1$ (lower curve) and ${\cal K}_2$ in units of $M^2$ as functions of
$|p|/M$ plotted for 
(a) $\theta=0$ and (b) $\theta=\pi/2$.}   
\label{f7}
\end{figure}

In order to further illustrate the physics of this kind of spontaneous
symmetry breaking let us look at the spectrum of small fluctuations $\delta
\chi$ (particle excitations) about the ground state. Inserting $\chi=
\chi_{\rm min}+\delta\chi$ into (\ref{a1}) yields
\begin{eqnarray}
\label{b7}
S\left[\chi_{\rm min}+\delta\chi\right]=S\left[\chi_{\rm min}\right]
+S_{\rm fluct}\left[\chi_{\rm min},\delta\chi\right]+{\cal O}\left(\delta
\chi^3\right)
\end{eqnarray}
with
\begin{eqnarray}
\label{b8}
S_{\rm fluct}\left[\chi_{\rm min},\delta\chi\right]
=\frac{1}{2}\int d^dx\,\left(\delta\chi^*,\delta\chi\right)\,
\hat{S}^{(2)}[\chi_{\rm min}]\,\left(\begin{array}{c}\delta\chi\\  \delta
\chi^*\end{array}\right)
\end{eqnarray}
where $\hat{S}^{(2)}$ is the $2\times 2$ matrix differential operator 
corresponding to the
second functional derivatives of $S$ with respect to $\chi$ and $\chi^*$. In
appendix \ref{D} we diagonalize this operator by a linear transformation
from $(\delta\chi,\delta\chi^*)$ to new, real fields $\Phi_1$ and $\Phi_2$. In
terms of the new fields $S_{\rm fluct}$ reads, in momentum representation,
\begin{eqnarray}
\label{b9}
S_{\rm fluct}\left[\chi_{\rm min},\delta\chi\right]
=\frac{1}{2}\int\frac{d^dp}{(2\pi)^d}\,\left\{\Phi_1(-p)\,
{\cal K}_1\left(p^2,\theta\right)\,\Phi_1(p)
+\Phi_2(-p)\,{\cal K}_2\left(p^2,\theta\right)\,\Phi_2(p)\right\}
\end{eqnarray}
where $\theta$ is the angle between $n_\mu$ and $p_\mu$. The kinetic terms
${\cal K}_{1/2}$ are given by
\begin{eqnarray}
\label{b10}
{\cal K}_1\left(p^2,\theta\right)&=&M^2+4\cos^2\theta\,p^2+\frac{p^4}{M^2}-
\sqrt{M^4+16\frac{\cos^2\theta}{M^2}\,p^6}
=4\cos^2\theta\,p^2+{\cal O}(p^4)\nonumber\\
{\cal K}_2\left(p^2,\theta\right)&=&M^2+4\cos^2\theta\,p^2+\frac{p^4}{M^2}+
\sqrt{M^4+16\frac{\cos^2\theta}{M^2}\,p^6}\nonumber\\
&=&2M^2+4\cos^2\theta\,p^2+{\cal O}(p^4)=4\cos^2\theta
\left(\frac{M^2}{2\cos^2\theta}+p^2\right)+{\cal O}(p^4)\,.    
\end{eqnarray}
In FIG. \ref{f7} we plot ${\cal K}_1$ and ${\cal K}_2$ for 
$\theta=0$ and
$\theta=\pi/2$ which amounts to excitations propagating parallel and
perpendicular to the vacuum direction $n_\mu$, respectively. For other
values of $\theta$ the behaviour is qualitatively similar.

The mode $\Phi_1$ is a massless excitation with an inverse propagator
${\cal K}_1$ which vanishes at $p=0$ for any $\theta$. It represents the
Goldstone boson of the spontaneously broken global {\sf U(1)} symmetry. The
spontaneous breaking of the {\sf SO($d$)} rotation symmetry manifests itself 
in the $\theta$-dependence of the propagator. In particular, one should note
that momenta orthogonal to $Q_{0\mu}=M n_\mu$ (i.e. $\cos\theta=0$) correspond
to the ``Goldstone direction'' with ${\cal K}_1=p^4/M^2$. 
The mode $\Phi_2$ is massive for all values of $\theta$. Its 
direction-dependent mass is given by $M/(\sqrt{2}\cos\theta)$.
It is the analogue of the ``Higgs'' or ``radial'' mode in the more familiar
case of a spontaneous symmetry breaking by a Mexican-hat potential.
\renewcommand{\baselinestretch}{1}
\small\normalsize
\renewcommand{\baselinestretch}{1.5}
\small\normalsize
\section{The effective average action}
\label{3}
Before embarking on the detailed analysis of the theory let us briefly comment
on the method we are going to employ. In order to find the true vacuum state
we need nonperturbative information about the (conventional) effective action
$\Gamma[\phi]$ where $\phi$ is the vacuum expectation value of $\chi$. As we 
mentioned in the introduction we regard $\Gamma$ as the physical limit of the 
effective average action $\Gamma_k[\phi]$ which was introduced in refs. 
\cite{avact}. The functional $\Gamma_k$ results from the
classical action $S$ by integrating out only the field modes with momenta
larger than the infrared cutoff $k$. Changing $k$ corresponds to a Wilson type
renormalization. The conventional effective action $\Gamma$
is recovered in the limit $k\rightarrow 0$. In the space of all actions,
the renormalization group trajectory $\Gamma_k$, $k\in[0,\infty)$,
interpolates between the classical action $S=\Gamma_{k\rightarrow\infty}$ and
the standard effective action $\Gamma=\Gamma_{k\rightarrow 0}$. It can be
obtained by solving an exact functional renormalization group
equation.

The infrared cutoff is implemented by modifying the path integral for the
generating functional of the connected Green's functions according to
\begin{eqnarray}
\label{i3}
\exp\left\{W_k\left[J\right]\right\}&=&\int{\cal D}\chi\,
\exp\bigg\{-S\left[\chi\right]-\int d^dx\,\chi^*(x)\,{\cal R}_k\left(
-\Box\right)\,\chi(x)\nonumber\\
& &+\int d^dx\,\left\{\chi^*(x)\,J(x)+J^*(x)\,\chi(x)\right\}
\bigg\}\,.
\end{eqnarray}
Here ${\cal R}_k(p^2)$ is a to some extent arbitrary positive function which
interpolates smoothly between ${\cal R}_k(p^2)\rightarrow k^2$ for $p^2
\rightarrow 0$ and ${\cal R}_k(p^2)\rightarrow 0$ for $p^2\rightarrow\infty$.
It suppresses the contribution of the small-momentum modes by a mass term
$\propto k^2$ which acts as the IR cutoff. In practical computations
${\cal R}_k(p^2)=p^2\left[\exp\left(p^2/k^2\right)-1\right]^{-1}$ has been
used often. In cases where this does not lead to UV divergences the condition
${\cal R}_k(p^2)\rightarrow 0$ for $p^2\rightarrow\infty$ can be relaxed and
one may also use a constant cutoff function ${\cal R}_k=k^2$ which amounts to
a momentum independent mass term. For the purposes of our present
investigation this simple cutoff will be sufficient.

The Legendre transform of $W_k$ reads
\begin{eqnarray}
\label{i4}
\widetilde{\Gamma}_k\left[\phi\right]=\int d^dx\,\left\{\phi^*\,J+J^*\,
\phi\right\}-W_k\left[J\right]
\end{eqnarray}
where the functional $J=J(\phi)$ is obtained by inverting the relations
\begin{eqnarray}
\label{i5}
\phi(x)=\frac{\delta W_k}{\delta J^*(x)}\;,\;\;\phi^*(x)=\frac{\delta W_k}
{\delta J(x)}
\end{eqnarray}
which define the $k$-dependent average field $\phi=\left<\chi\right>$. The 
effective average action is obtained from (\ref{i4}) by subtracting the cutoff 
term at the level of the average fields:
\begin{eqnarray}
\label{i6}
\Gamma_k\left[\phi\right]=\widetilde{\Gamma}_k\left[\phi\right]
-\int d^dx\,\phi^*\,{\cal R}_k\left(-\Box\right)\,\phi
\end{eqnarray}
It can be represented by an implicit functional integral
\begin{eqnarray}
\label{-a8}
\exp\left\{-\Gamma_k\left[\phi\right]\right\}&=&\int{\cal D}\sigma\,
\exp\Bigg\{-S[\phi+\sigma]-\int d^dx\,
\sigma^*\,{\cal R}_k\,\sigma\nonumber\\
& &+\int d^dx\,\left(\sigma^*\,\frac{\delta\Gamma_k}{\delta
\phi^*}[\phi]+\sigma\,\frac{\delta\Gamma_k}{\delta
\phi}[\phi]\right)\Bigg\}\,.
\end{eqnarray}
Here we introduced the shifted field
\begin{eqnarray}
\label{-a8.5}
\sigma\equiv\chi-\phi\,.
\end{eqnarray}
It has been shown \cite{avact} that the functional defined in this way has the
interpolating pro\-per\-ties stated above, and that it satisfies an exact
renormalization group equation. In the present paper we shall not use this
flow equation but rather calculate $\Gamma_k$ directly from the above
definition. We shall evaluate the path integral (\ref{-a8}) or (\ref{i3}) by 
means of a
saddle point approximation. It will turn out that the global minimum of the
Euclidean action can be found analytically. By expanding about this minimum
and properly taking its degeneracies into account we will be able to deduce
the essential features of the true vacuum already from the lowest order of the
saddle point approximation. In the following we shall disregard loop effects.
Because of the nontrivial vacuum structure of the theory, already the
``tree-level'' approximation yields nontrivial renormalization effects.

For configurations $\phi_{\rm sw}$ close to the spin wave solution (\ref{w0})
the saddle point approximation for the effective average action obeys for all
$k$
\begin{eqnarray}
\label{m3}
\Gamma_k\left[\phi_{\rm sw}\right]=S\left[\phi_{\rm sw}\right]\,.
\end{eqnarray}
This follows directly from (\ref{-a8}) by an expansion of $\phi_{\rm sw}$ 
around (\ref{w0}) and an expansion in powers of $\sigma$. The terms linear in
$\sigma$ cancel and the term quadratic in $\sigma$ is positive by virtue of 
the discussion in the last section. If the solution (\ref{w0}) corresponds to 
the absolute minimum of $\Gamma_k$ for small positive $k$ the rotation 
symmetry is spontaneously broken. Indeed, if one adds a small source term 
$J(\vec{q})$ the degeneracy of the minimum is lifted and the absolute
minimum will occur for the same momentum direction and the same phase as $J$.
In the limit $J\rightarrow 0$ the spin-wave configuration and its orientation
persist - this is spontaneous symmetry breaking. We will see below that this
spontaneous symmetry breaking of rotation symmetry is indeed realized for our
scalar model.
\section{Effective average action near the origin and classical 
renormalization effects}
\label{10}
In order to establish that the spin-wave configuration (\ref{w0}) corresponds
to the absolute minimum of $\Gamma_k$ for small positive $k$ we need to
understand the behaviour of $\Gamma_k$ for arbitrary $\phi$. In this section
we concentrate on small values of $\phi$, i.e. configurations that are far 
away from the spin-wave solution. In particular, we want to understand the
issue of convexity of the effective action for $k\rightarrow 0$ in case of
spontaneous breaking of rotation symmetry.

We are interested in the expansion of $\Gamma_k$ quadratic in $\phi$, i.e. the
two-point function at the origin. For this purpose we exploit the fact that the
origin $\phi=0$ always corresponds to a stationary point of
$\Gamma_k$ by virtue of the symmetries. By continuity we infer that - except 
for the phase with spontaneous symmetry breaking - small 
values of $\phi$ correspond to small sources $J$. In this case we will 
directly evaluate the scale dependent generating functional $W_k$ for the 
theory (\ref{a1}) which is given by the path integral
\begin{eqnarray}
\label{a53}
\exp\left\{W_k\left[J\right]\right\}
=\int {\cal D}\chi\,\exp\left\{-S_k^J[\chi;J]
\right\}
\end{eqnarray} 
with
\begin{eqnarray}
\label{a5}
S_k^J[\chi;J]&=&S_k\left[\chi\right]
-\int d^dx\,\left\{J^*\,\chi+J\,\chi^*\right\}
\end{eqnarray}
and
\begin{eqnarray}
\label{a0}
S_k\left[\chi\right]
=\int d^dx\,\left\{\chi^*\,\omega_k(-\Box)\,\chi
+\frac{\lambda}{2}\,|\chi|^4\right\}\,.
\end{eqnarray}
Here
\begin{eqnarray}
\label{-a1}
\omega_k(-\Box)\equiv\Omega(-\Box)+{\cal R}_k(-\Box)
\end{eqnarray}
is the complete kinetic operator. In momentum space it reads
\begin{eqnarray}
\label{-a2}
\omega_k(p^2)=-p^2+{\cal R}_k(p^2)+\frac{(p^2)^2}{2M^2}\;.
\end{eqnarray}
For values of the cutoff $k$ which are larger than some critical value 
$k_{\rm cr}$ 
the modified inverse propagator $\omega_k(p^2)$ is positive for any value of
the momentum $p$ since the regulator term ${\cal R}_k(p^2)$ overrides any
negative contribution which could come from $\Omega(p^2)$. Every admissible
function ${\cal R}_k(p^2)$ leads to a $k_{\rm cr}$ which equals $M$ times a 
number
of order unity. The mass-type cutoff ${\cal R}_k=k^2$ yields $k_{\rm cr}=M/
\sqrt{2}$, for instance. As a consequence, for large values of $k$, and in
particular for $k$ close to the UV cutoff $\Lambda$ where we start the
renormalization group evolution, there is no instability. As we lower the
cutoff $k$ below the critical value, $\omega_k(p^2)$ becomes negative for
certain modes. Obviously the modes with momenta in a band centered at
$p^2=M^2$ become unstable first. Finally, at $k=0$, all modes with
$p^2$ in the interval $(0,2M^2)$ have become unstable in the sense that
by exciting such modes we can push the value of the Euclidean action below
its value for the trivial configuration $\chi(x)=0$. We suspect that this
instability causes the modes with typical momenta of the order of $M$ to
``condense'', and it is this phenomenon which we are going to investigate.

The functionals $W_k[J]$ and $\Gamma_k[\phi]$ enjoy the same invariance
properties as the classical action. In addition to Poincar\'{e} symmetry
they are invariant under $J\rightarrow J\exp(i\varphi)$
and $\phi\rightarrow \phi\exp(i\varphi)$, respectively. This implies that
the series expansion of $W_k$ in powers of the sources, provided that it 
exists, contains only terms with an equal number of $J$ and $J^*$. In 
particular, the quadratic term displaying the effective propagator reads
\begin{eqnarray}
\label{-a4}
W_k^{\rm quad}\left[J\right]=\int d^dx\,d^dy\,J^*(x)\,G_k(x,y)\,J(y)
\equiv \int d^dx\,J^*(x)\,G_k(-\Box)\,J(x)
\end{eqnarray}
so that
\begin{eqnarray}
\label{b0}
W_k\left[J\right]=W_k\left[0\right]+W_k^{\rm quad}\left[J\right]+\ldots
\end{eqnarray}
where the dots stand for terms of order $J^2{J^*}^2$. Likewise,
Legendre-transforming (\ref{-a4}) and using (\ref{i6}) leads to
\begin{eqnarray}
\label{b1}
\Gamma_k\left[\phi\right]=-W_k\left[0\right]+\Gamma_k^{\rm quad}\left[\phi
\right]+{\cal O}\left(\phi^2\,\phi^{*2}\right)
\end{eqnarray}
where
\begin{eqnarray}
\label{-a5}
\Gamma_k^{\rm quad}\left[\phi\right]=\int d^dx\,d^dy\,\phi^*(x)\,\Sigma_k(x,y)
\,\phi(y)\equiv\int d^dx\,\phi^*(x)\,\Sigma_k(-\Box)\,\phi(x)
\end{eqnarray}
with
\begin{eqnarray}
\label{-a6}
\Sigma_k(-\Box)=G_k(-\Box)^{-1}-{\cal R}_k(-\Box)\;.
\end{eqnarray}
Henceforth we interpret $\Sigma_k$ as the differential operator
$\Sigma_k(-\Box)$ or as the function $\Sigma_k(p^2)$ in momentum space.

Frequently we shall evaluate the effective average action for plane-wave
configurations $\phi(x)=A\,\exp(ip_\mu x_\mu+i\alpha)$. Then 
{\sf U(1)}-invariance leads to 
\begin{eqnarray}
\label{w1} 
\Gamma_k[A\,\exp(ip_\mu x_\mu+i\alpha)]\equiv {\cal V}\, U_k(A;p^2)\;\;\;\;\; 
\left({\cal V}\equiv \int d^dx\right)
\end{eqnarray}
where we
refer to $U_k(A;p^2)$ as the effective potential for the mode with
momentum $p$. Note that while $\Gamma_k$ gives rise to an infinity of such
``effective potentials'', one for each momentum, it is clear that the totality
of all those potentials contains much less information than $\Gamma_k$ since
the correlations between different momentum modes are not specified.

In the subsequent sections we evaluate the functional integral (\ref{a53}) by
means of a saddle point approximation and determine $\Gamma_k$ directly from
its definition (\ref{i4})-(\ref{i6}) rather than by solving a flow equation.
Denoting the global minimum of $S_k^J$ by $\chi_{\rm min}(J)$ the lowest order
approximation reads
\begin{eqnarray}
\label{-a7}
W_k\left[J\right]``=\mbox{''}-S_k^J\left[\chi_{\rm min}(J);J\right]\,.
\end{eqnarray}
Under certain conditions (for instance for $J=0$ and $k$ sufficiently small) 
the global minimum is 
degenerate. In this case eq. (\ref{-a7}) is only symbolic (indicated by the
equality sign ``='') and one should sum over the degenerate minima. We ignore
this subtlety for a moment.

We shall see that already the lowest order approximation (\ref{-a7}) 
encapsulates all the essential physics which leads to a dynamical 
stabilization. At first sight this might seem surprising because one could
suspect that (\ref{-a7}) leads to the trivial result $\Gamma_k=S$. In fact, if
one performs a conventional perturbative loop expansion in a theory with a 
positive definite Hessian $S^{(2)}\equiv\delta^2 S/\delta\chi\,\delta\chi^*$ 
one has the standard lowest-order results $\Gamma=S$ and $\chi_{\rm min}=
\phi$. Since all contributions from fluctuations (one-loop
determinant etc.) are neglected in (\ref{-a7}) one might wonder how, 
nevertheless, nontrivial renormalization effects can occur. 

This point can be understood by looking at the integro-differential equation 
for $\Gamma_k$ given by eq. (\ref{-a8}). Now we try to find the 
($k$-dependent) global minimum $\sigma_{\rm min}\equiv\sigma_{\rm min}(\phi)$ 
of the complete action in the exponential of (\ref{-a8}). It satisfies
\begin{eqnarray}
\label{-a9}
\frac{\delta S}{\delta\phi^*}\left[\phi+\sigma_{\rm min}\right]+{\cal R}_k\,
\sigma_{\rm min}=\frac{\delta\Gamma_k}{\delta\phi^*}\left[\phi\right]\;.
\end{eqnarray}
For very large values of $k$, the term $\int d^dx\;\sigma^*{\cal R}_k\sigma$
in (\ref{-a8})
strongly suppresses fluctuations with $\sigma\neq 0$, so the main contributions
to the integral (\ref{-a8}) will come from small oscillations about 
$\sigma_{\rm min}=0$, which is indeed the global minimum in this case.
In this case one has $\chi_{\rm min}\equiv\sigma_{\rm min}+\phi=\phi$ and 
(\ref{-a9}) is obeyed for $\Gamma_k=S$. When we lower $k$ towards zero 
it often happens that $\sigma_{\rm min}=0$ continues to be
the absolute minimum of the action. This is the case for classically stable
theories with a positive definite Hessian $S^{(2)}$ where no condensation 
phenomena occur.

On the other hand, eq. (\ref{-a9}) may admit also nontrivial solutions with
$\sigma_{\rm min}\neq 0$ for $k$ small enough. They are relevant in the case of
instabilities where the Hessian $S^{(2)}[\phi]$ develops negative eigenvalues.
Setting
$\sigma=\sigma_{\rm min}+\psi$ and expanding up to second order in $\psi$ one
has, with $\chi_{\rm min}\equiv\phi+\sigma_{\rm min}$,
\begin{eqnarray}
\label{-a10}
\exp\left\{-\Gamma_k\left[\phi\right]\right\}&=&\exp\Bigg\{
-S\left[\chi_{\rm min}\right]-\int d^dx\,\sigma_{\rm min}^*\,
{\cal R}_k\,\sigma_{\rm min}\nonumber\\
& &+\int d^dx\,\left(\sigma_{\rm min}^*\,
\frac{\delta\Gamma_k}{\delta\phi^*}\left[\phi\right]+\sigma_{\rm min}\,
\frac{\delta\Gamma_k}{\delta\phi}\left[\phi\right]\right)\Bigg\}
\nonumber\\
& &\times\int{\cal D}\psi\,\exp\left\{-\int d^dx\,\psi^*
\left(S^{(2)}
\left[\chi_{\rm min}\right]+{\cal R}_k\right)\psi\right\}\,.
\end{eqnarray}
For the ``correct'' saddle point $S^{(2)}[\chi_{\rm min}]+{\cal R}_k$ should 
be positive semidefinite. The zero modes in case of degeneracy of the minimum 
lead to an integration over the vacuum manifold. In lowest order one neglects 
the remaining Gauss integral involving the positive eigenvalues of $S^{(2)}[
\chi_{\rm min}]$ and finds
\begin{eqnarray}
\label{-a11}
\Gamma_k\left[\phi\right]&``=\mbox{''}&S\left[\chi_{\rm min}
\right]+\int d^dx\,\sigma_{\rm min}^*\,{\cal R}_k\,\sigma_{\rm min}
-\int d^dx\,\left(\sigma_{\rm min}^*\,
\frac{\delta\Gamma_k}{\delta\phi^*}\left[\phi\right]+\sigma_{\rm min}\,
\frac{\delta\Gamma_k}{\delta\phi}\left[\phi\right]\right)
\end{eqnarray}
For $\sigma_{\rm min}\neq 0$ (i.e. if $\chi_{\rm min}\neq\phi$) this is still
a complicated differential equation for $\Gamma_k$ whose solution is {\it not}
given by $\Gamma_k=S$. Hence the lowest order (classical) term of the saddle
point approximation does indeed give rise to nontrivial renormalization effects
leading to $\Gamma_k\neq S$. In fact, all the qualitative features of the
condensation phenomena we are interested in are described by this classical 
term alone. The higher order corrections yield minor quantitative corrections
only. For this to be the case it is important to correctly identify the 
absolute minimum $\chi_{\rm min}$ of the (total) action. At next-to-leading 
order the fluctuations $\psi$ modify the r.h.s. of (\ref{-a11}) by a one-loop 
term $\ln{\rm Det}'\left(S^{(2)}[\chi_{\rm min}]+{\cal R}_k\right)$ where the
prime at the determinant indicates that the zero modes of $S^{(2)}[\chi_{\rm
min}]$ are excluded. For our case of interest this correction is typically 
small, at least for $d>2$.

In ref. \cite{RW90} this formalism has been applied to the familiar spontaneous
symmetry breaking by a Mexican-hat potential. It was found that the classical
term of the saddle point expansion describes all salient features of the 
effective potential such as the approach of convexity for $k\rightarrow 0$,
and that the one-loop determinant does not modify its qualitative properties.
In ref. \cite{ABP99} similar classical instability driven renormalization 
effects have been found in the framework of the Wegner-Houghton equation. For
an investigation of a different scalar theory with a higher derivative
kinetic term containing the usual positive quadratic kinetic term see ref. 
\cite{BMP}.
\section{A sufficient condition for the absolute minimum}
\label{4}
Later on  we shall use the saddle point approximation to calculate the 
effective average action. This requires the knowledge of the absolute minimum
of the action $S_k^J$ (\ref{a5}). In order to identify the absolute minimum, 
we use a method similar to the one outlined in \cite{RW90}. First we split the
field $\chi$ into a classical solution that minimizes the action and an 
arbitrary, not necessarily small deformation 
$\delta\chi$:
\begin{eqnarray}
\label{a6}
\chi(x)=\chi_{\rm min}(x)+\delta\chi(x)
\end{eqnarray}
If we insert this ansatz into the action (\ref{a5}) and take advantage of the
fact that $\chi_{\rm min}$ is a solution of the equation of motion (e.o.m.) 
\begin{eqnarray}
\label{a10}
\frac{\delta S_k^J}{\delta\chi^*}&=&0\Leftrightarrow
\left[\Omega(-\Box)+{\cal R}_k(-\Box)+\lambda\,\left|\chi\right|^2\right]\chi
=J
\end{eqnarray}
we obtain
\begin{eqnarray}
\label{a11}
S_k^J\left[\chi;J\right]=
S^J_k\left[\chi_{\rm min};J\right]
+\Delta S_k\left[\chi_{\rm min},\delta\chi\right]
\end{eqnarray}
where $\Delta S_k$ is defined as
\begin{eqnarray}
\label{a12}
\lefteqn{\Delta S_k\left[\chi_{\rm min},\delta\chi\right]
\equiv\int d^dx\,\bigg\{
\delta\chi^*\left[\Omega(-\Box)+{\cal R}_k(-\Box)\right]\delta\chi
+\frac{\lambda}{2}\bigg[\chi_{\rm min}^2\,\left(\delta\chi^*\right)^2}
\nonumber\\
& &+{\chi_{\rm min}^*}^2\,\left(\delta\chi\right)^2
+4\left|\chi_{\rm min}\right|^2\,\left|\delta\chi\right|^2
+2\chi_{\rm min}\,\delta\chi^*\,\left|\delta\chi\right|^2
+2\chi_{\rm min}^*\,\delta\chi\,\left|\delta\chi\right|^2
+\left|\delta\chi\right|^4\bigg]\bigg\}\,.
\end{eqnarray}
One can read off immediately that a solution of the e.o.m. corresponds to the 
absolute minimum
of $S_k^J$ if 
\begin{eqnarray}
\label{a13}
\Delta S_k\left[\chi_{\rm min},\delta\chi\right]\ge 0
\end{eqnarray}
is satisfied for any deformation $\delta\chi$.
Because $\Delta S_k$ does not explicitly contain the sources this condition 
is the same for $S^J_k$ and $S^{J=0}_k=S_k$.

It is useful to rewrite the expression for $\Delta S_k$ in the following 
manner.
\begin{eqnarray}
\label{a15}
\Delta S_k\left[\chi_{\rm min},\delta\chi\right]
&=&\int d^dx\,
\delta\chi^*\left[\Omega(-\Box)+{\cal R}_k(-\Box)
+\lambda\,\left|\chi_{\rm min}\right|^2\right]\delta\chi\nonumber\\
& &+\frac{\lambda}{2}\int d^dx\,
\left(\chi_{\rm min}\,\delta\chi^*+\chi_{\rm min}^*\,\delta\chi
+\left|\delta\chi\right|^2\right)^2
\end{eqnarray}
Because $\chi_{\rm min}\,\delta\chi^*+\chi_{\rm min}^*\,\delta\chi
=2{\rm Re}\left(\chi_{\rm min}\,\delta\chi^*\right)$ is real
the integrand of the second integral in eq. (\ref{a15}) cannot become 
negative. 
Therefore we may conclude that 
\begin{eqnarray}
\label{a100}
\int d^dx\,
\delta\chi^*\left[\Omega(-\Box)+{\cal R}_k(-\Box)
+\lambda\,\left|\chi_{\rm min}\right|^2\right]\delta\chi\ge 0
\end{eqnarray} 
for any $\delta\chi(x)$ is a sufficient condition for $\chi_{\rm min}$
being the absolute minimum of the action $S_k^J$ for some suitable source $J$
given by the solution of eq. (\ref{a10}).
\section{Searching for the absolute minimum}
\label{5}
The aim of this section is to find the field configuration that minimizes the 
action $S_k^J$ {\it globally}. Here we start with the case of vanishing 
external sources and relegate the case $J\neq 0$ to appendix \ref{A}.
From now on we use the mass-type cutoff function ${\cal R}_k=k^2$. It 
simplifies the algebra and allows for particularly transparent results.

We are searching for the minimum of $S_k$, so we have to solve the
e.o.m.
\begin{eqnarray}
\label{a18}
& &\left[\omega_k(-\Box)+\lambda\,\left|\chi\right|^2\right]\chi=0\;,\;\;
\omega_k\equiv\Omega+k^2=\Box+\frac{\Box^2}{2M^2}+k^2\;.
\end{eqnarray}
Its most obvious solution is the one of vanishing $\chi$, $\chi_{\rm min}=0$.
We can now employ the tool we developed in the last section to find out whether
$\chi_{\rm min}=0$ constitutes the global minimum. First we insert this 
solution into eq. (\ref{a15}) which yields 
\begin{eqnarray}
\label{a19}
\Delta S_k\left[\chi=0,\delta\chi\right]=\int d^dx\,\delta\chi^*
\left[\Omega(-\Box)+k^2\right]\delta\chi
+\frac{\lambda}{2}\int d^dx\,\left|\delta\chi\right|^4\,.
\end{eqnarray}
The next step is to use the Fourier representation
\begin{eqnarray}
\label{27}
\delta\chi(x)=\int\frac{d^dp}{(2\pi)^d}\,\widetilde{\delta\chi}(p)\,\exp\left(
ip_\mu x_\mu\right)
\end{eqnarray}
and to go over to momentum space for the first term:
\begin{eqnarray}
\label{a28}
\Delta S_k\left[\chi=0,\delta\chi\right]
=\int\frac{d^dp}{(2\pi)^d}\,\left|\widetilde{\delta\chi}(p)\right|^2
\,\left[\Omega(p^2)+k^2\right]
+\frac{\lambda}{2}\int d^dx\,\left|\delta\chi\right|^4
\end{eqnarray}
For $k^2\ge M^2/2$ one finds that $\omega_k(p^2)=\Omega(p^2)+k^2\ge 0$ and, as
a consequence, $\Delta S_k$ is always positive (or zero). In this range of $k$
$\chi_{\rm min}=0$ represents the
field configuration that minimizes the action globally. However, if $k^2<
M^2/2$, then, for certain deformations with a sufficiently small amplitude,
$\Delta S_k$ becomes negative. In that case $\chi_{\rm min}=0$ cannot be the
absolute minimum. 

In order to find the global minimum for $k^2<M^2/2$ we try 
the plane wave ansatz
\begin{eqnarray}
\label{a20}
\chi_{\rm min}(x)=\chi_0\,\exp\left(iQ_\mu x_\mu+i\alpha\right)\;,\;\;
\chi_0,\,\alpha\;
{\rm real}.
\end{eqnarray}
Inserting ({\ref{a20}}) into the e.o.m. leads to the condition
\begin{eqnarray}
\label{a21}
\chi_0(Q^2)=\sqrt{-\omega_k(Q^2)/\lambda}\,.
\end{eqnarray}
Obviously $\chi_0(Q^2)$ is real only if $\omega_k(Q^2)\le 0$. 

It will be useful to calculate the action for the solution
$\chi_{\rm min}$ determined above. We obtain
\begin{eqnarray}
\label{a22}
S_k\left[\chi=\chi_0(Q^2)\,\exp\left(i Q_\mu x_\mu+i\alpha\right)\right]
=-\frac{{\cal V}}{2\lambda}\,\omega_k^2(Q^2)\,.
\end{eqnarray}
The action (\ref{a22}) still depends on a free\footnote{$Q^2$
is free only within certain bounds, because we chose $\chi_0$ to be real.} 
parameter, the momentum squared $Q^2$. 
Hoping to find the absolute minimum for the 
region of small $k$-values by using the plane wave ansatz, we minimize the 
expression (\ref{a22}) with respect to $Q^2$. This leads to the condition
$ Q^2=M^2$. Thus we have found a candidate for the global minimum in the 
region $k^2\le M^2/2$ which is given by
\begin{eqnarray}
\label{a25}
\chi_{\rm min}(x)=\hat{\chi}_0\,\exp\left(i Q_{0\mu}x_\mu+i\alpha\right)\;;\;\;
Q_{0\mu}\equiv M\,n_\mu\,.
\end{eqnarray}
Here $n_\mu$ is an arbitrary unit vector ($n^2=1$), the phase $\alpha$
is taken to be in $[0,2\pi)$
and
\begin{eqnarray}
\label{a25,5}
\hat{\chi}_0\equiv\sqrt{-\frac{1}{\lambda}\omega_k(M^2)}
=\sqrt{\frac{1}{\lambda}\left(\frac{1}{2}M^2-k^2\right)}\;;\;\;k^2\le
\frac{1}{2}M^2\,.
\end{eqnarray}

By inserting the solution (\ref{a25}) into $\Delta S_k$ (\ref{a15}) we can 
check whether it corresponds to the global minimum.
In momentum space, the l.h.s. of (\ref{a100}) takes the form
\begin{eqnarray}
\label{a26}
\int\frac{d^dp}{(2\pi)^d}\,
\left|\widetilde{\delta\chi}(p)\right|^2\,\left[\Omega(p^2)+\frac{1}{2}M^2
\right]\,.
\end{eqnarray}
Because of $\Omega(p^2)\ge -M^2/2$, this integral is nonnegative and thus the
sufficient condition for a global minimum is fulfilled. 

Eq. (\ref{a25}) actually describes a whole family of degenerate minima
parametrized by the phase $\alpha$ and the unit vector $n_\mu$. As a 
consequence, the ``vacuum manifold'' is $S^1\times S^{d-1}$.

We conclude that the nontrivial solution $(\ref{a25})$ corresponds to the 
absolute minimum of $S_k$ for the 
region $k^2\le M^2/2$, while the solution $\chi_{\rm min}=0$ constitutes the 
absolute minimum for $k^2\ge M^2/2$.
The two solutions coincide at $k^2=M^2/2$.
\section{Renormalization group flow of the zero field propagator}
\label{6}
In this section we compute the scale dependent 2-point function $\Sigma_k$
(inverse propagator) which was introduced in eq. (\ref{-a5}). The saddle point
expansion about the global minimum of $S_k^J$ leads to
\begin{eqnarray}
\label{a57}
\exp\left\{W_k\left[J\right]\right\}&=&
N_k\,\sum\limits_{\chi_{\rm min}}\exp
\bigg\{-S_k^J\left[\chi_{\rm min};J\right]
-\frac{1}{2}\,\ln{\rm Det}'\hat{S}_k^{(2)}\left[\chi_{\rm min}\right]
+\ldots\bigg\}\,.
\end{eqnarray}
Here $\sum_{\chi_{\rm min}}$ denotes a
symbolic summation or integration over the possibly degenerate absolute
minima. The first term inside the curly brackets of (\ref{a57}) represents the
dominant classical term. The second one contains the one-loop effects, and it 
will be neglected in the following. The normalization constant $N_k$ depends
on the nature of the minimum and its degeneracy. It will be adjusted such that
$W_k[0]$ is continuous with the initial condition $W_{k\rightarrow\infty}[0]
=0$.

Because $\chi_{\rm min}(x)$ is now the minimum of the action $S_k^J$ in  
presence of a source it becomes here  
a functional of the source $J(x)$. The determination of $\chi_{\rm min}=
\chi_{\rm min}(J)$ for an arbitrary source would be a formidable task. The 
fact that we only want to study the bilinear term in $W_k$ simplifies the
problem considerably, since in this case the minimum is needed only for an 
infinitesimal source $J$, and $\chi_{\rm min}(J)$ can be obtained by a 
perturbative expansion about the source-free minimum $\chi_{\rm min}(J=0)\equiv
\varphi^{(0)}$ given by eq. (\ref{a25}). We write $J(x)\equiv\varepsilon j(x)$ 
where $\varepsilon$ counts the powers of $J$ (with $j$ taken to be of order 
$\varepsilon^0$) and we expand the saddle point according to
\begin{eqnarray}
\label{a61}
\chi_{\rm min}(x)=\varphi^{(0)}(x)+\varepsilon\,\varphi^{(1)}(x)+\varepsilon^2
\,\varphi^{(2)}(x)+\ldots
\end{eqnarray}
In order to compute the $J^*J$-term in $W_k$ it is sufficient to know the first
order correction $\varphi^{(1)}$. It satisfies the linearized equation of 
motion
\begin{eqnarray}
\label{a69}
\left(\begin{array}{c}J\\ J^*\end{array}\right)
=\varepsilon\hat{S}_k^{(2)}\left[\varphi^{(0)}\right]\,
\left(\begin{array}{c}\varphi^{(1)}\\ \varphi^{(1)*}\end{array}\right)\;.
\end{eqnarray}
The operator $\hat{S}_k^{(2)}$ is defined as
\begin{eqnarray}
\label{a55}
\hat{S}_k^{(2)}\left[\chi\right]\,\delta^d(x-y)=
\left(\begin{array}{cc}\frac{\delta^2 S_k}{\delta\chi^*(x)\,\delta\chi(y)}
& \frac{\delta^2 S_k}{\delta\chi^*(x)\,\delta\chi^*(y)}\\
\frac{\delta^2 S_k}{\delta\chi(x)\,\delta\chi(y)}
& \frac{\delta^2 S_k}{\delta\chi(x)\,\delta\chi^*(y)}\end{array}\right)
\end{eqnarray}
which yields
\begin{eqnarray}
\label{a55,5}
\hat{S}_k^{(2)}\left[\chi\right]
\equiv\left(\begin{array}{cc}\omega_k(-\Box)+2\lambda\,\left|\chi\right|^2
& \lambda\,\chi^2\\
\lambda\,\chi^{*2}
& \omega_k(-\Box)+2\lambda\,\left|\chi\right|^2\end{array}\right)\,.
\end{eqnarray}
For $\chi=\varphi^{(0)}$ this matrix operator is nonsingular except for the 
case where ``$-\Box=M^2$'' which corresponds to the vacuum degeneracy for $k^2
\le M^2/2$. For sources $J(p^2\neq Q_0^2)$ one can obtain 
$\varphi^{(1)}$ by inverting (\ref{a69}).
This leads to the following expansion of $S_k^J$ up to terms quadratic in the
sources:
\begin{eqnarray}
\label{a66}
\lefteqn{S_k^J\left[\chi_{\rm min};J\right]
=S_k^J\left[\varphi^{(0)}+\varepsilon\,\varphi^{(1)}+\ldots;J\right]}
\nonumber\\
&=&S_k^J\left[\varphi^{(0)};J\right]
-\varepsilon\int d^dx\,\left\{J\,\varphi^{(1)*}+J^*\,\varphi^{(1)}\right\}
\nonumber\\
& &+\frac{1}{2}\,\varepsilon^2\,\int d^dx\,
\left(\varphi^{(1)*},
\varphi^{(1)}\right)\,
\hat{S}^{(2)}_k\left[\varphi^{(0)}\right]\,
\left(\begin{array}{c}
\varphi^{(1)}\\ \varphi^{(1)*}\end{array}\right)
+{\cal O}\left(\varepsilon^3\right)\nonumber\\
&=&S_k^J\left[\varphi^{(0)};J\right]
-\frac{1}{2}\,\int d^dx\,\left(J^*,J\right)\,
\hat{S}^{(2)}_k\left[\varphi^{(0)}\right]^{-1}\,
\left(\begin{array}{c}
J\\ J^*\end{array}\right)
+{\cal O}\left(J^2{J^*}^2\right)
\end{eqnarray}
From now on no reference to the $J$-dependence of $\chi_{\rm min}$ is made
any more, and knowledge of the source-free saddle point $\varphi^{(0)}$ is
sufficient in order to determine $G_k$ from (\ref{a57}).

Let us first look at the upper region of the renormalization group evolution
where $k^2\ge k_{\rm cr}^2=M^2/2$. Then $\chi_{\rm min}=\varphi^{(0)}\equiv 0$
is the relevant nondegenerate global minimum, and from (\ref{a57}) with
(\ref{a66}) we obtain
\begin{eqnarray}
\label{-a12}
W_k\left[J\right]=\int d^dx\,J^*\,\omega_k(-\Box)^{-1}\,J+\ldots
\end{eqnarray}
($W_k[0]=0$ is achieved by setting $N_k=1$ for all $k\ge k_{\rm cr}$.)
Hence $G_k$ equals the tree level (cutoff) propagator so that by eqs.
(\ref{-a5}) and (\ref{-a6})
\begin{eqnarray}
\label{-a13}
\Sigma_k(p^2)=\Omega(p^2)\;,\;\;\forall k^2\ge M^2/2\;.
\end{eqnarray}
We conclude that during the early stage of the evolution, as long as $k$ is
larger than $k_{\rm cr}$, the inverse propagator is not renormalized, (except
for the loop effects neglected here).

The situation changes once $k$ drops below $k_{\rm cr}$. Then the relevant 
saddle point is given by eq. (\ref{a25}) which represents a degenerate minimum
parametrized by the unit vector $n_\mu$ and the phase $\alpha$. Hence the
``summation'' over $\chi_{\rm min}$ in eq. (\ref{a57}) amounts to an
integration over the vacuum manifold $S^1\times S^{d-1}$. This integration
will turn out crucial in order to restore {\sf U(1)}- and 
Poincar\'{e}-invariance. (In appendix \ref{D} we comment on the 
situation where the integration over the vacuum manifold is omitted.)
In eq. (\ref{a66}) we also need
\begin{eqnarray}
\label{-a14}
S_k^J\left[\varphi^{(0)};J\right]&=&
S_k\left[\varphi^{(0)}\right]
-\int d^dx\left\{J\,\varphi^{(0)*}+J^*\,\varphi^{(0)}\right\}\nonumber\\
&=&-\frac{{\cal V}}{2\lambda}\,\omega_k^2(M^2)
-\sqrt{-\frac{1}{\lambda}\omega_k(M^2)}\,\left(\tilde{J}(Q_0)\,e^{-i\alpha}
+\tilde{J}^*(Q_0)\,e^{i\alpha}
\right)
\end{eqnarray}
where $\tilde{J}(p)$ denotes the Fourier-transform of $J(x)$. For source 
functions $J(x)$ which do not contain Fourier components with
$p^2=Q_0^2=M^2$ the last term of (\ref{-a14}) drops out. Upon
expanding also the l.h.s. of eq. (\ref{a57}) the latter boils down to\footnote{
The limits $J\rightarrow 0$ and ${\cal V}\rightarrow\infty$ may not commute. 
Here we take the infinite volume limit at the end. Obviously the ``correct''
order of limits depends on the physical situation one has in mind. In ordinary
quantum field theories or in statistical mechanics the limits are usually
performed in reverse order, i.e. first ${\cal V}\rightarrow\infty$ and then
$J\rightarrow 0$. But since we would like to use the present scalar result as
a guideline for Euclidean quantum gravity on spacetimes with a finite volume 
($S^d$, etc.) we have to perform the limit  $J\rightarrow 0$ first.} 
\begin{eqnarray}
\label{-a15}
\exp\left\{W_k[0]\right\}\left(1+W_k^{\rm quad}\left[J\right]+\ldots\right)
&=&N_k\,\exp\left\{\frac{{\cal V}}{2\lambda}\,\omega_k^2(M^2)\right\}\left.
\int\limits_0^{2\pi} d\alpha\int\limits_{S^{d-1}}d\mu(n)\right\{1+\nonumber\\
& &\left.+\frac{1}{2}\,\int d^dx\,\left(J^*,J\right)\,
\hat{S}^{(2)}_k\left[\varphi^{(0)}\right]^{-1}\,
\left(\begin{array}{c}
J\\ J^*\end{array}\right)+\ldots\right\}
\end{eqnarray}
Here $\int d\mu(n)$ denotes the {\sf SO($d$)}-invariant measure on $S^{d-1}$. 
Since $\omega_{k_{\rm cr}}(M^2)=0$, we see that $W_k[0]$ is continuous at 
$k=k_{\rm cr}\equiv M/\sqrt{2}$ if we set
\begin{eqnarray}
\label{-a16}
N_k=\left[2\pi\,\Omega_{d-1}\right]^{-1}
\end{eqnarray}
for $k<k_{\rm cr}$ where $\Omega_{d-1}\equiv\int_{S^{d-1}}d\mu(n)$ is the
volume of the $(d-1)$-sphere. Thus we are left with
\begin{eqnarray}
\label{b2}
W_k\left[0\right]=-C_k\,{\cal V}
\end{eqnarray}
where
\begin{eqnarray}
\label{-b3}
C_k=-\frac{1}{2\lambda}\omega_k^2(M^2)=-\frac{1}{8\lambda}\left(M^2-2k^2
\right)^2\,.
\end{eqnarray}
and
\begin{eqnarray}
\label{-a17}
W_k^{\rm quad}\left[J\right]&=&\left[2\pi\,\Omega_{d-1}\right]^{-1}
\int\limits_0^{2\pi}
d\alpha\int\limits_{S^{d-1}}d\mu(n)\,\frac{1}{2}\int d^dx\,\left(J^*,J\right)\,
\hat{S}^{(2)}_k\left[\varphi^{(0)}\right]^{-1}\,
\left(\begin{array}{c}
J\\ J^*\end{array}\right)\,.
\end{eqnarray}
Before we can perform the integration over the vacuum manifold in the above 
expression we need to know the inverse of the matrix operator $\hat{S}_k^{(2)}
[\varphi^{(0)}]$ (\ref{a55,5}). This inverse is found to be given by
\begin{eqnarray}
\label{a105}
\lefteqn{\hat{S}^{(2)}_k\left[\varphi^{(0)}\right]^{-1}}
\nonumber\\
&=&\left(\begin{array}{cc}\left[\omega_k(-{\cal D}^{*2})-2\,\omega_k(M^2)
\right]\,P_k(\Box,{\cal D}^{*2}) & 
\omega_k(M^2)\,e^{2i M n_\mu x_\mu+2i\alpha}\,P_k(\Box,{\cal D}^2)\\
\omega_k(M^2)\,e^{-2i M n_\mu x_\mu-2i\alpha}\,P_k(\Box,{\cal D}^{*2}) 
& \left[\omega_k(-{\cal D}^2)-2\,\omega_k(M^2)\right]
\,P_k(\Box,{\cal D}^2)\end{array}\right)
\end{eqnarray}
where
\begin{eqnarray}
\label{a106}
P_k(\Box,{\cal D}^2)
=\left[\left[\omega_k(-{\cal D}^2)-2\,\omega_k(M^2)\right]
\left[\omega_k(-\Box)-2\omega_k(M^2)\right]-\omega_k^2(M^2)\right]^{-1}
\end{eqnarray}
and
\begin{eqnarray}
\label{a107}
{\cal D}_\mu=\partial_\mu+2iM n_\mu\;,\;\;{\cal D}_\mu^*=\partial_\mu-2i 
M n_\mu\,.
\end{eqnarray}
After inserting the expression (\ref{a105}) into $\rm eq.\,(\ref{-a17})$ we 
turn our attention to the integration over $\alpha$ first. Due to this 
integration the off-diagonal entries of the matrix operator
$\hat{S}^{(2)}_k[\varphi^{(0)}]^{-1}$ 
yield vanishing contributions. This removes the $J^2$- and $J^{*2}$-terms and 
thus guarantees the {\sf U(1)}-invariance. 
For the diagonal entries which are independent of the phase $\alpha$ the
effect of this integration is just a factor of $2\pi$.

Next we go over to momentum space and then interchange the two remaining (up 
to this point, independent) integrations over the momentum $p$ and the 
direction of the unit vector $n$ such that the latter is performed first. The
easiest way to solve this integral is to choose the $n$-coordinate system in 
such a way that one of its axes is parallel to the momentum $p$. After 
introducing polar coordinates for the integration over 
the $(d-1)$-sphere we have to deal with only one nontrivial angular
integral since apart from the volume element of the $(d-1)$-sphere the
integrand just depends on the angle $\theta$ enclosed by
$p$ and $n$ which enters via the scalar product $p_\mu n_\mu=|p|
\cos\theta$. The remaining $d-2$ angular integrals amount to the volume of
a $(d-2)$-sphere and thus to a factor $\Omega_{d-2}$. This leads to
\begin{eqnarray}
\label{-a18}
W_k\left[J\right]=W_k[0]+
\int\frac{d^dp}{(2\pi)^d}\,\tilde{J}^*(p)\,\tilde{G}_k(|p|)\,
\tilde{J}(p)
+{\cal O}\left(J^2{J^*}^2\right)
\end{eqnarray}
where ($\kappa\equiv k/M$)
\begin{eqnarray}
\label{-a19}
\lefteqn{\tilde{G}_{k=M\kappa}\Big(|p|=Mq\Big)=M^{-2}\frac{\Omega_{d-2}}
{\Omega_{d-1}}\int\limits_0^{(1+\delta_{d,2})\pi}d\theta\,\sin^{d-2}\theta
\left[\left(\frac{\left(q^2-1\right)^2}{2}+\frac{1}{2}-\kappa^2\right)\right.}
\nonumber\\
& &\times\left.\left(\frac{q^4}{2}-4q^3\cos\theta+q^2\left(3
+8\cos^2\theta\right)-12q\cos\theta+5-\kappa^2\right)
-\frac{1}{4}+\kappa^2-\kappa^4\right]^{-1}\nonumber\\
& &\times\left(\frac{q^4}{2}-4q^3\cos\theta+q^2\left(3
+8\cos^2\theta\right)-12q\cos\theta+5-\kappa^2\right)\,.
\end{eqnarray}
Here we can identify $G_k(p^2)\equiv\tilde{G}_k\left(|p|\equiv (p_\mu 
p_\mu)^{1/2}\right)$ as the propagator defined in eq. (\ref{-a4}).

In summary the effective average action for small $\phi$ is given by
\begin{eqnarray}
\label{-a20}
\Gamma_k\left[\phi\right]=-\frac{{\cal V}}{8\lambda}\left(M^2-2k^2\right)^2
+\int\frac{d^dp}{(2\pi)^d}\,\tilde{\phi}^*(p)\,\tilde{\Sigma}_k(|p|)\,
\tilde{\phi}(p)+{\cal O}\left(\phi^2{\phi^*}^2\right)
\end{eqnarray}
such that the bilinear or kinetic term $\Sigma_k(p^2)\equiv\tilde{\Sigma}_k
\left(|p|\equiv (p_\mu p_\mu)^{1/2}\right)$ is obtained as 
$\tilde{\Sigma}_k(|p|)\equiv\tilde{G}_k(|p|)^{-1}-k^2$.

We have calculated the remaining integral over $\theta$ for $d=2$, $3$ and $4$
and found explicit analytic expressions for the kinetic term at arbitrary 
values of $k$. The expressions for $d=3$, $k=0$ and $d=4$, $k=0$  can be found
in appendix \ref{C}. We omitted the one for $d=2$ dimensions and those for 
$k>0$ because the formulas are extremely lengthy.
\renewcommand{\baselinestretch}{1}
\small\normalsize
\begin{figure}[ht]   
\begin{minipage}{6cm}
        \epsfxsize=6cm
        \epsfysize=5cm
        \centerline{\epsffile{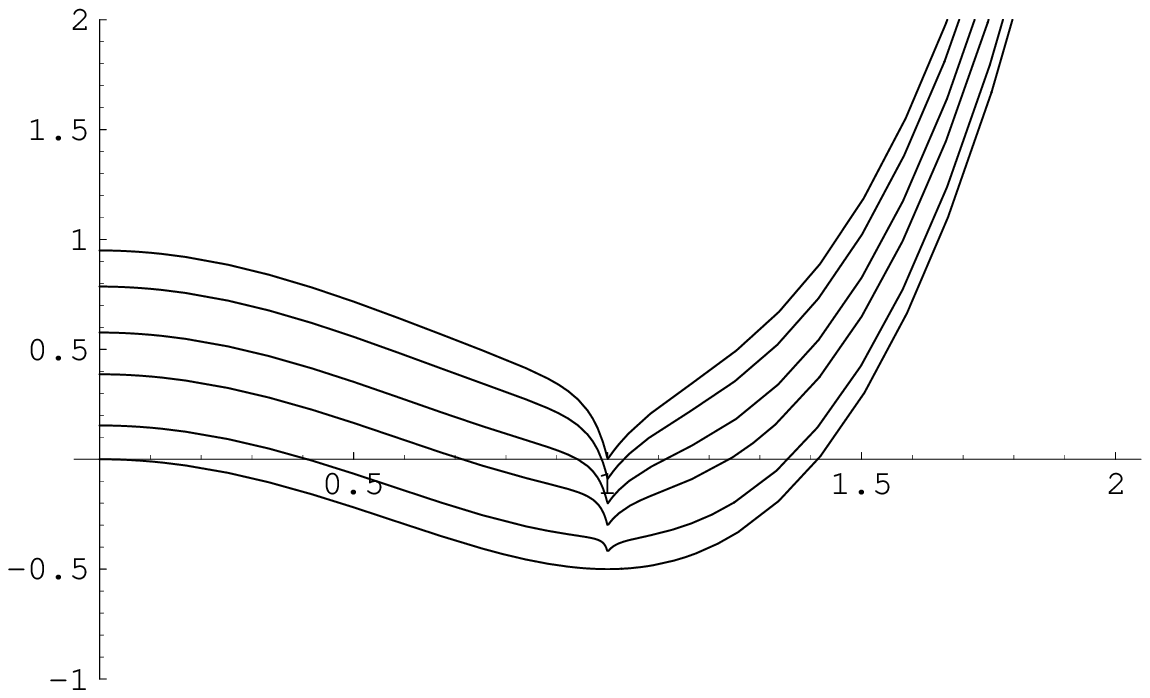}}
\centerline{(a)}
\end{minipage}
\hfill
\begin{minipage}{6cm}
        \epsfxsize=6cm
        \epsfysize=5cm
        \centerline{\epsffile{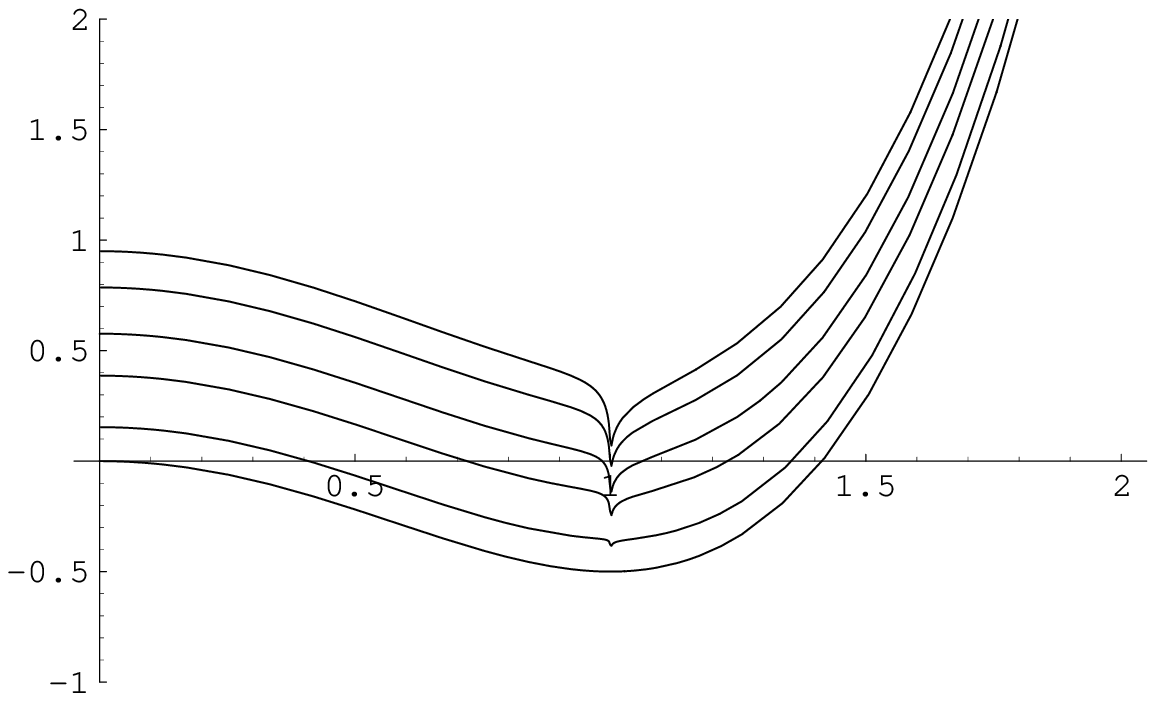}}
\centerline{(b)}
\end{minipage}
\vspace{0.3cm}
\caption{Kinetic term $\tilde{\Sigma}_k$ (in units of $M^2$) in (a)
$d=3$ and (b) $d=4$ dimensions as a function of $|p|/M$ plotted for distinct
$k$-values. The lowest curve corresponds to $\kappa\equiv k/M=k_{\rm cr}/M
=1/\sqrt{2}$ 
whereas the remaining curves correspond to $\kappa$-values 0.65, 0.55, 0.45, 
0.3 and 0 in increasing order.}
\label{f1}
\end{figure}
\begin{figure}[ht]
        \epsfxsize=12cm
        \epsfysize=10cm
        \centerline{\epsffile{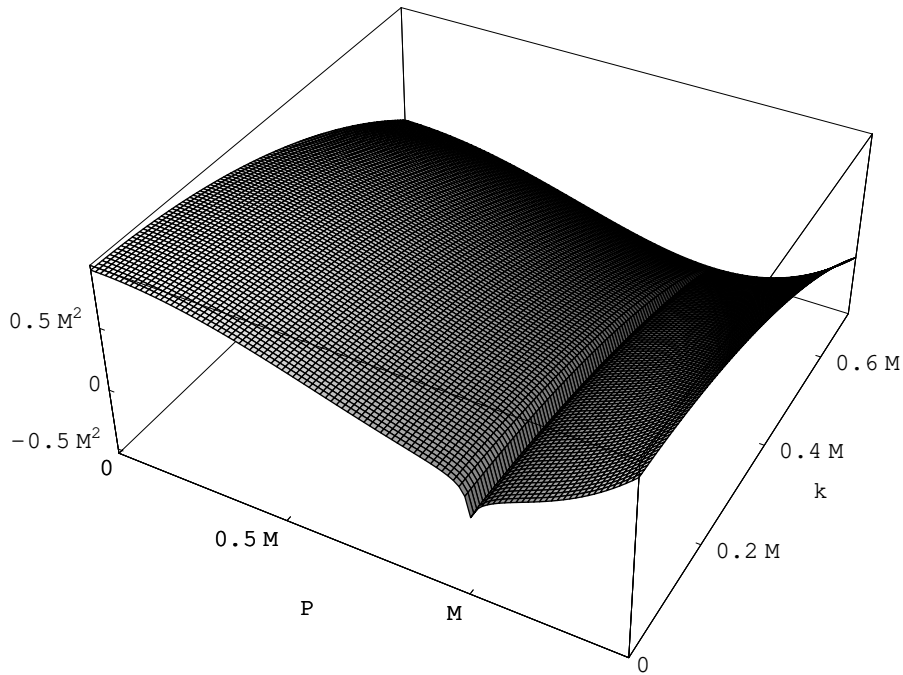}}
\caption{%
Kinetic term $\tilde{\Sigma}_k$ in $d=4$ dimensions as a 
function of ${\rm p}\equiv |p|$  and the cutoff $k$.}
\label{f6}
\end{figure}
\renewcommand{\baselinestretch}{1.5}
\small\normalsize

The behaviour of the various kinetic terms is illustrated in figures 
\ref{f1} and \ref{f6} by means of two different kinds of plots. In the
3D-plot the kinetic term is presented as a function of the two variables
$|p|$ and $k$ whereas the 2D-plots show certain slices of the corresponding
3D-plots at distinct $k$-values. (We have displayed only one of the 
3D-plots here.) Since for $k\ge M/\sqrt{2}$ the kinetic term is not
renormalized and thus does not flow at all, each plot starts at 
$k=M/\sqrt{2}\equiv k_{\rm cr}$.

Obviously the plots reflect the same qualitative behaviour for the values of
$d$ considered here. At $k=k_{\rm cr}$ the kinetic term is given by 
$\Omega(p^2)$ which is characterized by the features discussed above. As we 
move on towards smaller scales $k$ the whole function is lifted so that the 
$p$-region where the kinetic term $\tilde{\Sigma}_k(|p|)$ takes negative values
shrinks more and more. Finally, at $k=0$, $\tilde{\Sigma}_k(|p|)$ has become
completely nonnegative rendering the theory stable. In the course of this
evolution the kinetic term builds up a downward peak at $|p|=M$ which grows 
sharper while the theory is evolved towards smaller values of $k$. At the
point $|p|=M$, $\tilde{\Sigma}_{k=0}(|p|)$ has a cusp-type singularity at 
which it is nondifferentiable but still continuous. It is the shape of this 
cusp that makes up the main difference between the results for $d=2$ (which we
have determined but not displayed here), $d=3$ and $d=4$. Obviously it turns 
sharper as we increase the dimensionality.
The dressed inverse propagator $\Omega_{\rm eff}\equiv\Sigma_{k=0}$ had been 
presented in the introduction already (FIG. \ref{f13}).
\section{Regions in field space without classical renormalization}
\label{7}
The results for the effective average action derived so far concentrated on the
structure of its kinetic term for small $\phi$. In the following sections we 
discuss the essential features of the complete effective average action and 
of the potential for our model. 

In section \ref{10} it was pointed out that there are regions in $\phi$-space
where no classical renormalization occurs and thus
\begin{eqnarray}
\label{d1}
\Gamma_k\left[\phi\right]=S[\phi]=\int d^dx\left\{\phi^*\,\Omega(-\Box)\,\phi
+\frac{\lambda}{2}\left|\phi\right|^4\right\}\,.
\end{eqnarray}
Obviously we obtain $\Gamma_k=S$ for the whole $\phi$-space if 
$k\ge k_{\rm cr}$, since then $\hat{S}_k^{(2)}[\chi]$ is positive definite 
for any $\chi$.

Let us now consider the case $k<k_{\rm cr}$. It follows immediately from the
discussion in section \ref{10} that no classical renormalization is expected
for those $\phi$ which differ only slightly from $\chi_{\rm min}$.
Furthermore, it is easy to see that as long as $|\phi|$ is
sufficiently large we have $\chi_{\rm min}=\phi$ in leading order and thus 
$\Gamma_k=S$. To show this we assume that 
\begin{eqnarray}
\label{e0}
J=\lambda\,\left|\chi_{\rm min}\right|^2\,\chi_{\rm min}
\end{eqnarray}
provides a good approximation to the e.o.m. (\ref{a10}) which means that 
$\chi_{\rm min}$, and thus $J$, are large enough to render the 
$\omega_k(-\Box)$-term negligible. For sources leading to (\ref{e0}) the 
generating functional $W_k$ is then approximately given by
\begin{eqnarray}
\label{e1}
W_k[J]=-S_k^J\left[\chi_{\rm min};J\right]=-\int d^dx\left\{\frac{\lambda}{2}\,
\left|\chi_{\rm min}\right|^4-J\,\chi_{\rm min}^*-J^*\,\chi_{\rm min}\right\}
\end{eqnarray}
which coincides with the Legendre-transform of
\begin{eqnarray}
\label{e2}
\widetilde{\Gamma}_k[\phi]=\int d^dx\,\frac{\lambda}{2}\left|\phi\right|^4
=S[\phi]
\end{eqnarray}
where $\phi=\chi_{\rm min}$.
If $\int d^dx\,\phi^*\omega_k(-\Box)\phi$ is at least of the same order of 
magnitude as $\int d^dx\, k^2|\phi|^2$ this implies $\Gamma_k=S$. Otherwise
we have to assume in addition that $k^2|\phi|^2<<\lambda|\phi|^4$ in order to
obtain the same result $\Gamma_k=S$. 
Obviously, in both cases it is the large field $\phi=\chi_{\rm min}$ which 
causes the Hessian $\hat{S}_k^{(2)}[\chi_{\rm min}]$ to be positive definite,
thereby acting as a cutoff.

In order to make this argument more precise we 
restrict our considerations from now on to the subspace of plane-wave fields
\begin{eqnarray}
\label{d2}
\phi(x)=A\,\exp(ip_\mu x_\mu+i\beta) 
\end{eqnarray}
where $A$ is a real, positive amplitude. We start from the assumption that
indeed 
\begin{eqnarray}
\label{e3}
\Gamma_k\left[A\,\exp(ip_\mu x_\mu+i\beta)\right]=S\left[A\,\exp(ip_\mu x_\mu
+i\beta)\right]\,.
\end{eqnarray}
Then the results obtained in appendix \ref{A} may be applied to
(\ref{e3}) in order to determine conditions on the parameter $A$ which tell
us when $\Gamma_k=S$ is satisfied.

Inserting (\ref{d2}) into (\ref{d1}) and adding the cutoff term 
$k^2\int d^dx\,|\phi|^2$ leads to
\begin{eqnarray}
\label{d3}
\widetilde{\Gamma}_k\left[A\,\exp(ip_\mu x_\mu+i\beta)\right]
\equiv {\cal V}\left[U_k(A;p^2)+k^2\,A^2\right]
={\cal V}\left[\omega_k(p^2)\,A^2+\frac{\lambda}{2}\,A^4\right]
\end{eqnarray}
from which we obtain
\begin{eqnarray}
\label{d4}
J=\frac{\delta\widetilde{\Gamma}_k}{\delta\phi^*}\left[A\,\exp\left(ip_\mu 
x_\mu+i\beta\right)\right]=\left[\omega_k(p^2)\,A+\lambda\,A^3\right]
\,\exp\left(ip_\mu x_\mu+i\beta\right)
\end{eqnarray}
By setting $J=\varepsilon\,\exp(ip_\mu x_\mu+i\beta)$ this equation takes the
form
\begin{eqnarray}
\label{d5}
\omega_k(p^2)\,A+\lambda\,A^3=\varepsilon
\end{eqnarray}
which is equivalent to eq. (\ref{a33}) with $\hat{\chi}_0$ replaced with $A$. 
Thus we can read off the solution of (\ref{d5}) from appendix \ref{A} which 
is given by
\begin{eqnarray}
\label{d6}
A(\varepsilon;p)=\left\{\begin{array}{ll}\chi_0(\varepsilon;p)\,,\; & 
\omega_k(p^2)
\ge 0\\\tilde{\chi}_0(\varepsilon;p)\,,\; & \omega_k(p^2)<0\end{array}\right.
\end{eqnarray}
where $\chi_0(\varepsilon;p)$ and $\tilde{\chi}_0(\varepsilon;p)$ are defined 
by eqs. (\ref{a34}), (\ref{a38,5}), respectively.\footnote{In case of
$\omega_k(p^2)<0$ we only consider the solution which is continuous at 
$\Delta=0$ (see eq. (\ref{a118})) and omit the other two branches.}
Eqs. (\ref{d5}), (\ref{d6}) may now be used to perform the
Legendre-transformation from $\widetilde{\Gamma}_k$ to $W_k$ which yields
\begin{eqnarray}
\label{e4}
W_k\left[J=\varepsilon\,\exp(ip_\mu x_\mu+i\beta)\right]
&=&-{\cal V}\left[\omega_k(p^2)
\,A^2(\varepsilon;p)+\frac{\lambda}{2}\,A^4(\varepsilon;p)-2\varepsilon\,
A(\varepsilon;p)\right]\nonumber\\
&=&-S_k^J\left[A(\varepsilon;p)\,\exp(ip_\mu x_\mu+i\beta);
\varepsilon\,\exp(ip_\mu x_\mu+i\beta)\right]\,.
\end{eqnarray}
According to appendices \ref{A}, \ref{A1} this expression equals 
$-S_k^J[\chi_{\rm min};J]$ if and only if either (I): $k\ge k_{\rm cr}$ or 
(II): $k<k_{\rm cr}$ and $A(\varepsilon;p)\ge A_{\rm triv}$. Here
\begin{eqnarray}
\label{z0}
A_{\rm triv}\equiv\hat{\chi}_0=\sqrt{\frac{1}{2\lambda}(M^2-2k^2)}
\end{eqnarray}
denotes the critical value of the amplitude beyond which $\Gamma_k$ becomes
trivial, i.e. equal to $S$.\footnote{ 
If $k<k_{\rm cr}$ and $|p|=M$, $A=A_{\rm triv}$ is assumed for $\varepsilon=0$ 
which means that, in this special case, we have to insert the free, degenerate 
minimum (\ref{a25}) into $S_k^J$ in eq. (\ref{e4}). However, the result for 
$S_k^J[\chi_{\rm min};J]$ is not influenced by the degeneracy of this 
solution.} 
It depends on $k$ but is independent of the momentum $p$. At the initial point
$k=k_{\rm cr}$ we have $A_{\rm triv}=0$, but when we lower $k$ $A_{\rm triv}$
grows and the $A$-interval with nontrivial renormalization effects expands.

In terms of the effective potentials $U_k(A;p^2)$ for plane-wave fields which
we defined in eq. (\ref{w1}) the statement $\Gamma_k=S$ means
that $U_k(A;p^2)$ equals the classical potential
\begin{eqnarray}
\label{x0}
U^{(\rm cl)}(A;p^2)\equiv\Omega(p^2)\,A^2+\frac{\lambda}{2}\,A^4\,.
\end{eqnarray}
Its minimum is located at 
\begin{eqnarray}
\label{x1}
A_{\rm min}^{(\rm cl)}(p)=\sqrt{-\frac{1}{\lambda}\Omega(p^2)}
\end{eqnarray}
where it assumes the value
\begin{eqnarray}
\label{z2}
U^{(\rm cl)}(A_{\rm min}^{(\rm cl)}(p);p^2)=-\frac{\Omega^2(p^2)}{2\lambda}
\end{eqnarray}
For $k$ not too far below $k_{\rm cr}$, the classical minimum lies in the 
region with no renormalization ($A_{\rm triv}<A_{\rm min}^{(\rm cl)}$). But,
at $k=0$, $A_{\rm triv}$ is always larger than $A_{\rm min}^{(\rm cl)}$, except
for $p^2=M^2$ where $A_{\rm triv}=A_{\rm min}^{(\rm cl)}$.
\section{Structure of the effective average action}
\label{8}
In the previous section it was shown that the effective average action 
for plane-wave fields equals $S$ as long as $A\ge A_{\rm triv}(k)$, 
which will be refered to as the ``outer region'' of the effective 
potential. The inner region for $A<A_{\rm triv}(k)$ is characterized by
nontrivial instability induced classical renormalization. For 
$k\ge k_{\rm cr}$ no inner region occurs. For $k<k_{\rm cr}$ 
the determination of $U_k(A;p^2)$ in the inner region is more involved. 
Depending on the momentum $p$ several cases are to be distinguished.
\subsection{$|p|\neq M$}
First of all let us consider the case $|p|\neq M$. 
The definition of the effective potential
\begin{eqnarray}
\label{00}
\Gamma_k\left[A\,\exp(ip_\mu x_\mu+i\beta)\right]\equiv {\cal V}\,U_k(A;p^2)
\end{eqnarray}
corresponds to sources which are plane waves of the form $J=\varepsilon\,\exp
\left(ip_\mu x_\mu+i\beta\right)$. Here $\varepsilon$ is related to $A$, $p$ 
and $k$ via the source-field relation
\begin{eqnarray}
\label{e6}
\varepsilon=\frac{1}{2}\frac{\partial U_k(A;p^2)}{\partial A}
+k^2\,A\,.
\end{eqnarray}
Since according to section \ref{7} and appendix \ref{A} the outer region 
$A\ge A_{\rm triv}$ is already parametrized by the $\varepsilon$-values lying 
in the interval $[\varepsilon_k(p^2),\infty)$ the inner region of $U_k$ must 
correspond to $\varepsilon$-values in $[0,\varepsilon_k(p^2))$. The problem 
is that, in case of $\varepsilon<\varepsilon_k(p^2)$, we do not have an exact 
expression for the field configurations $\chi_{\rm min}$ which minimize the 
action $S_k^J$ globally. Thus we are not able to determine the inner region of
the potential $U_k(A;p^2)$ exactly. However, it is still possible to deduce 
the qualitative structure of $U_k(A;p^2)$ for $A<A_{\rm triv}$ by patching up
the available pieces of information. 

As in section \ref{6} we assume that, at least for sufficiently small 
$|\phi|$, $\Gamma_k$ is analytic in $\phi$ and $\phi^*$. Then it follows from 
{\sf U(1)}-invariance\footnote{Since the parameter $A$ is chosen to be real 
the {\sf U(1)}-invariance of $U_k$ is not manifest. It is nevertheless 
present, since $A$ has to be regarded as absolute value of a complex number.}
that $U_k(A;p^2)$ is an even function of $A$ and may be expanded about $A=0$ 
according to
\begin{eqnarray}
\label{e7}
U_k(A;p^2)=-\frac{1}{8\lambda}\left(M^2-2k^2\right)^2+\sum\limits_{n=1}^{
\infty}u^{(2n)}_k(p)\,A^{2n}\,.
\end{eqnarray}
In section \ref{6} we already calculated the first coefficient of this 
expansion $u^{(2)}_k(p)=\Sigma_k(p^2)$. 
In addition, an expression for $u^{(4)}_k$ is derived in appendix \ref{E}. 
It is given by
\begin{eqnarray}
\label{e8}
u^{(4)}_k(p)=-\left(\Sigma_k(p^2)+k^2\right)^4\,G^{(4)}_k(p^2)\,.
\end{eqnarray}
The 4-point function $G^{(4)}_k(p^2)$ has an extremely complicated structure; 
it has been evaluated numerically for some special values of $p$ and $k$ only.
Thus we know the behaviour of $U_k(A;p^2)$ for small values of $A$. 

On the other hand, we know that $U_k(A;p^2)=\Omega(p^2)\,A^2+(\lambda/2)\,A^4$
for $A\ge A_{\rm triv}$.

For the intermediate range of $A$-values where (\ref{e7}) is not applicable 
any more but $A$ is still below $A_{\rm triv}$ we have to find an 
interpolation which connects the small-$A$ region to the outer region 
$A\ge A_{\rm triv}$ in a qualitatively correct manner. We make the most natural
assumption that this interpolation is a minimal one in the sense that it leads
to as few as possible extrema of $U_k$. For instance, if $U_k(0;p^2)<U^{(\rm
cl)}(A_{\rm triv};p^2)$ and if the slope at $A=0$, i.e. $\Sigma_k$, is 
positive we interpolate with a monotonically increasing function from the 
small-$A$ region up to $A_{\rm triv}$. Likewise, if the slope at $A=0$ is 
negative the interpolating function is assumed to have a single minimum in the
intermediate region.

The justification for this scheme comes from three sources:
\begin{itemize}
\item[i)] The inclusion of the $A^4$-term confirms this picture in a part of
the parameter space.
\item[ii)] In the limit $k\rightarrow 0$ it leads to the expected convexity of
$U_0(A;p^2)$.
\item[iii)] The resulting $U_k(A;p^2)$ connects smoothly to $U_k(A;M^2)$
which can be evaluated exactly.
\end{itemize}
We shall discuss momenta $p$ with $\Omega(p^2)\ge 0$ and $\Omega(p^2)<0$
separately.
\subsubsection{The case $\Omega(p^2)\ge 0$}

\begin{figure}[ht]
\hbox to\hsize{\hss
       %\vspace{0.5cm}
        \epsfxsize=8cm
        \epsfysize=6cm
        \centerline{\epsffile{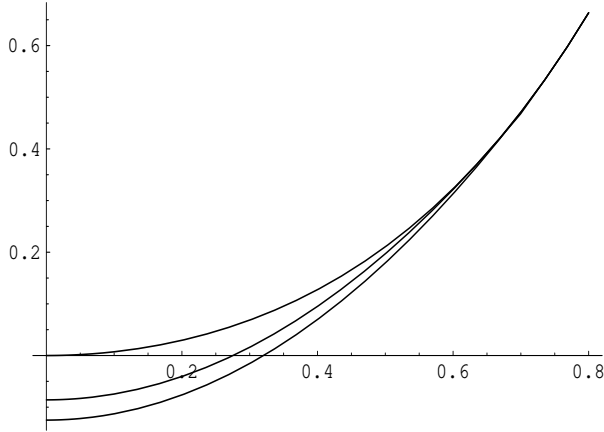}}\hss}
\caption{%
Qualitative scaling behaviour of $U_k(A;p^2)$ for $p^2\in\{0\}\cup [2M^2,
\infty)$, illustrated by three curves corresponding to the $k$-values 
$k=k_{\rm cr}$ (upper curve), $k=k_l$ where $0< k_l< k_{\rm cr}$ (curve in the
middle) and $k=0$.}
\label{f10}
\end{figure}

The case $\Omega(p^2)\ge 0$ corresponds to $p^2$-values lying in $\{0\}\cup 
[2M^2,\infty)$. In particular it includes the standard effective potential for
$p=0$. Since in this case $\Sigma_k> 0$ for all $k<k_{\rm cr}$ the minimal 
interpolation leads to the monotonically increasing function shown in FIG. 
\ref{f10}. The shape of this curve changes only insignificantly in the course 
of the evolution. The only change is a decrease of $C_k$ compensated
by an increase of the slope of the curve for $A<A_{\rm triv}$ such that (a) the
inner and the outer region join smoothly at $A=A_{\rm triv}$ and (b) $U_{k_2}
(A;p^2)<U_{k_1}(A;p^2)$ is valid for all $A<A_{\rm triv}$, if 
$k_2<k_1\le k_{\rm cr}$.
\renewcommand{\baselinestretch}{1}
\small\normalsize
\renewcommand{\baselinestretch}{1.5}
\small\normalsize
The inclusion of the $A^4$-term gives additional support to this picture.
It turns out that the $A^2$-term dominates the behaviour of $U_k$ for all
$A\le A_{\rm triv}$ if the difference between $|p|$ and $M$ is large enough. 
Then higher order terms like $u_k^{(4)}$ yield only minor corrections. 
\renewcommand{\baselinestretch}{1}
\small\normalsize
\begin{figure}[ht]
\begin{minipage}{6cm}
        \epsfxsize=6cm
        \epsfysize=5cm
        \centerline{\epsffile{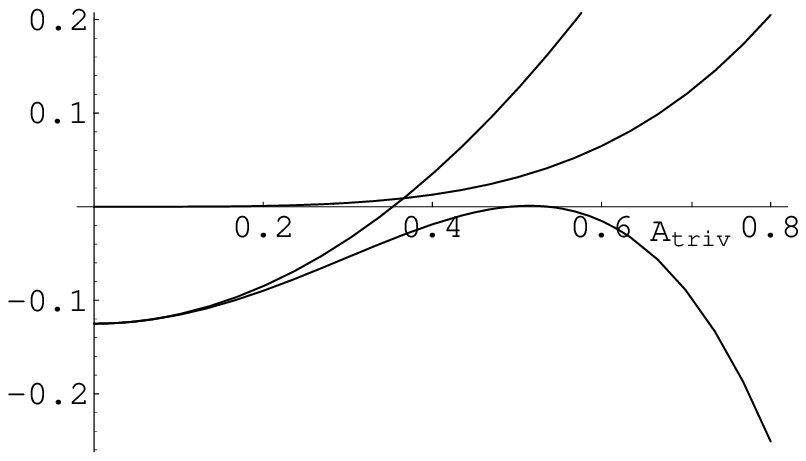}}
\centerline{(a)}
\end{minipage}
\hfill
\begin{minipage}{6cm}
        \epsfxsize=6cm
        \epsfysize=5cm
        \centerline{\epsffile{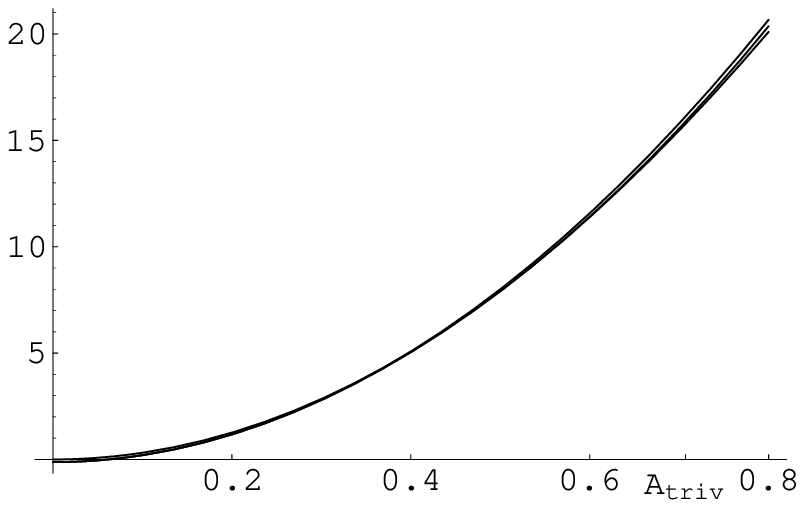}}
\centerline{(b)}
\end{minipage}
%\vskip 20pt
\vspace{0.3cm}
\caption{$U^{(\rm cl)}$, $U_{k=0}^{(2)}$ and $U_{k=0}^{(4)}$ (in units of 
$M^4/\lambda$) as functions of $A$ (in units of $M/\sqrt{\lambda}$).
(a): Case $p=0$. The curves for $U_{k=0}^{(2)}$ and $U_{k=0}^{(4)}$ are those
which coincide at $A=0$, where $U_{k=0}^{(4)}$ corresponds to the curve that
develops a negative slope for large $A$. The remaining curve
corresponds to $U^{(\rm cl)}$.
(b): Case $|p|=3M$. The difference between the three curves is hardly visible.
Near $A_{\rm triv}=M/\sqrt{2\lambda}$ we have $U_{k=0}^{(4)}<U^{(\rm cl)}<
U_{k=0}^{(2)}$.}
\label{f12}
\end{figure}
\renewcommand{\baselinestretch}{1.5}
\small\normalsize

In FIG. \ref{f12} we plotted $U^{(\rm cl)}(A;p^2)$ (which coincides with the
true $U_k$ for $k\ge k_{\rm cr}$, $A$ arbitrary and for $k<k_{\rm cr}$, $A\ge
A_{\rm triv}$) and the approximations
\begin{eqnarray}
\label{e10}
U_k^{(2)}(A;p^2)&\equiv&C_k+\Sigma_k(p^2)\,A^2\nonumber\\
U_k^{(4)}(A;p^2)&\equiv&C_k+\Sigma_k(p^2)\,A^2-\left(\Sigma_k(p^2)+k^2\right)^4
G_k^{(4)}(p^2)\,A^4
\end{eqnarray}
for $p=0$, $k=0$ and for $|p|=3M$, $k=0$. Obviously $p=0$ is not sufficiently
far away from $M$ because $U_0^{(2)}$ is a good approximation only for
very small $A$-values and then grows too fast to be able to merge with 
$U^{(\rm cl)}$ at $A_{\rm triv}$. The effect of the $A^4$-term is to bend this
curve downward. Hence $U_0^{(4)}$ is already fairly accurate for a larger 
range of $A$-values and it gets closer to the $U^{(\rm cl)}$-curve. For $A$
approaching $A_{\rm triv}$ also the $A^4$-approximation breaks down and it is
clear that higher orders are needed in order to hit the $U^{(\rm cl)}$-curve
at $A_{\rm triv}$.

For $|p|=3M$ the situation is much better. $U^{(\rm cl)}$, $U_0^{(2)}$ and
$U_0^{(4)}$ are virtually identical for all $A\le A_{\rm triv}=M/
\sqrt{2\lambda}$. (This is because $\Sigma_{k=0}(p^2)\approx\Omega(p^2)$ for 
large $p$.) Due to the corrections coming from the $A^4$-term we expect that
$U_0^{(4)}$ agrees better with the exact $U_0$ than $U_0^{(2)}$ but the 
difference between the two approximations is too tiny to be visible. 
However, we surely can infer from FIG. \ref{f12} that both $U_0^{(2)}$ and 
$U_0^{(4)}$ are convex.  
For even larger momenta the quality of the approximation (\ref{e7}) increases. 

From the general properties of Legendre transforms we expect 
$\lim_{k\rightarrow 0} U_k(A;p^2)$ to be a convex function of $A$, for all 
values of $p$. For the momenta considered here this convexity is indeed 
achieved, albeit in a somewhat trivial fashion since the potential was convex
from the outset.
\subsubsection{The case $\Omega(p^2)<0$}
Let us now turn to the case $\Omega(p^2)<0$, $|p|\neq M$, which is satisfied 
by all values of $p^2$ contained in $(0,2M^2)\setminus\{M^2\}$. For such 
momenta the $A^2$-expansion (\ref{e7}) yields no reliable results because
the higher order terms begin to dominate already at small values of $A<
A_{\rm triv}$. Those results are at most as reliable as the one for $p=0$, 
$k=0$ and they become increasingly worse as $|p|$ approaches $M$. Hence we 
have to apply a different method to determine the properties of $U_k$ in the 
inner region.

First of all it should be noted that in the course of the evolution 
$\Sigma_k(p^2)$ changes its sign which is obvious from the FIGS. 
\ref{f1} and \ref{f6}. There exists a scale $\hat{k}(p)<k_{\rm cr}$ at which
$\Sigma_{\hat{k}}(p^2)=0$; we have $\Sigma_k<0$ for $k>\hat{k}$ and $\Sigma_k
>0$ for $k<\hat{k}$.

\begin{figure}[ht]
\hbox to\hsize{\hss
        %\vspace{0.5cm}
        \epsfxsize=8cm
        \epsfysize=6cm
        \centerline{\epsffile{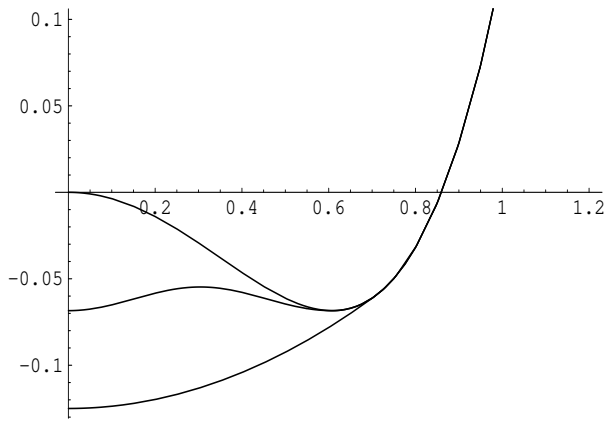}}\hss}
\caption{%
Qualitative scaling behaviour of $U_k(A;p^2)$ for $p^2\in(0,2M^2)\setminus 
\{M^2\}$, illustrated by three curves corresponding to the $k$-values 
$k=k_{\rm cr}$ (upper curve), $k=k_0$ where $0< k_0< k_{\rm cr}$ (curve in the
middle) and $k=0$.} 
\label{f9}
\end{figure}

Along with the evolution of $\Sigma_k$, i.e. the slope of $U_k$ at $A=0$, the
constant term $U_k(A=0;p^2)=C_k$ drops from $C_k=0$ at $k=k_{\rm cr}$ to
$C_k=-M^4/(8\lambda)$ at $k=0$. If one minimally interpolates between this
small-$A$ behaviour and $U^{(\rm cl)}$ one obtains the curves shown in 
FIG. \ref{f9}. Note that the change of sign of $\Sigma_k$ is crucial for 
achieving convexity in the limit $k\rightarrow 0$. As long as $k>\hat{k}(p)$, 
$U_k$ is not convex since the negative kinetic term $\Sigma_k$ causes $U_k$ to
decrease for very small values of $A$. 
\renewcommand{\baselinestretch}{1}
\small\normalsize
\renewcommand{\baselinestretch}{1.5}
\small\normalsize

In FIG. \ref{f9} we included an additional piece of information which is easy
to obtain. There exists a certain scale $k_0(p)\in(0,k_{\rm cr})$ at which
the value of the potential at $A=0$ equals its value at $A_{\rm triv}$:
$U_{k_0(p)}(0;p^2)=U_{k_0(p)}(A_{\rm triv};p^2)$. This means that $C_{k_0(p)}
=U^{(\rm cl)}(A_{\rm triv};p)$, and by eq. (\ref{-b3}) this condition is 
equivalent to $\Omega(p^2)=k_0^2-M^2/2$. Hence 
\begin{eqnarray}
\label{e12}
k_0(p)=\frac{1}{\sqrt{2}M}\left|M^2-p^2\right|\,.
\end{eqnarray}
Numerically we find that $k_0(p)<\hat{k}(p)$ for any value of $p$ considered
here. Therefore
the slope at the origin is positive for $k=k_0$. Since, at $A_{\rm triv}$,
$U^{(\rm cl)}$ has a positive slope too this means that $U_k$ has (at least)
one maximum and one minimum in the inner region. (The minimum is the expected
one, of course, essentially the minimum of $U^{(\rm cl)}$, corrected by the 
renormalization effects.) In order to obtain a convex potential in the limit $k
\rightarrow 0$ this local maximum must shrink as the theory is evolved towards
smaller $k$ until it vanishes at some scale between $k_0(p)$ and zero.

The results derived so far are sufficient to give a qualitative discription of
$U_k$ in the inner region. FIG. \ref{f9} illustrates its essential features as
well as the {\it exact} structure of the effective potential in the outer 
region. Again, the effective average potential becomes convex in the limit
$k\rightarrow 0$, this time in a less trivial fashion though. 
\subsection{$|p|=M$}
Let us finally  come to the ``resonant'' case $|p|=M$. It will turn out that
an {\it exact} expression for $U_k$ can be derived in this case. 

As in the case before we obtain from {\sf U(1)}-invariance
\begin{eqnarray}
\label{e14}
\Gamma_k\left[A\,\exp\left(iQ_{0\mu}x_\mu+i\beta\right)\right]\equiv
{\cal V}\,U_k(A;M^2)
\end{eqnarray}
where the corresponding sources are plane waves satisfying the source-field
relation (\ref{e6}) with $p$ replaced with $Q_0$. From section \ref{7} and
appendix \ref{A} we know that any source-amplitude $\varepsilon>0$ is related
to a field-amplitude $A\in(A_{\rm triv},\infty)$ via eq. (\ref{e6}) and that 
$\varepsilon=0$ yields $A_{\rm triv}$. Actually $\varepsilon=0$ corresponds 
to the {\it complete inner region} $A<A_{\rm triv}$ as well. To see this we 
have to look at the generating functional $W_k$ evaluated at $J=\varrho\,
\exp(iQ_{0\mu}x_\mu)$ where $\varrho\equiv\varepsilon\,\exp(i\beta)$ is 
complex. Inserting the global minimum $\chi_{\rm min}$ in presence of 
plane-wave sources, eq. (\ref{a38}), into $W_k[J]=-S_k^J[\chi_{\min}(J);J]$
and expanding this expression with respect to $|\varrho|\equiv\varepsilon$ 
about $|\varrho|=0$ 
yields 
\begin{eqnarray}
\label{e15}
\lefteqn{W_k\left[J=\varrho\,\exp(iQ_{0\mu}x_\mu)\right]}\nonumber\\
&=&{\cal V}\,\left[\frac{\omega_k^2(M^2)}{2\lambda}
+2\,|\varrho|\,\sqrt{-\frac{1}{\lambda}\omega_k(M^2)}
-\frac{|\varrho|^2}{2\,\omega_k(M^2)}\right]+{\cal O}(|\varrho|^3)\,.
\end{eqnarray}  
The crucial point is that the term linear in $|\varrho|$ causes 
$dW_k/d\varrho$ to be discontinuous at $\varrho=0$. 
In fact we obtain from 
(\ref{e15})
\begin{eqnarray}
\label{e16}
\lim\limits_{|\varrho|\rightarrow 0}\left[\frac{d}{d\varrho} W_k\left[\varrho\,
\exp(iQ_{0\mu}x_\mu)\right]\right]_{\varrho=|\varrho|\exp(i
\beta)}=\sqrt{-\frac{1}{\lambda}\omega_k(M^2)}\,\exp(-i\beta)
\equiv A_{\rm triv}\,\exp(-i\beta)
\end{eqnarray}
which shows that this derivative depends on $\beta$, i.e. on the direction in 
the complex plane from which $\rho=0$ is approached. 

\begin{figure}[ht]
\hbox to\hsize{\hss
        %\vspace{0.5cm}
        \epsfxsize=8cm
        \epsfysize=6cm
        \centerline{\epsffile{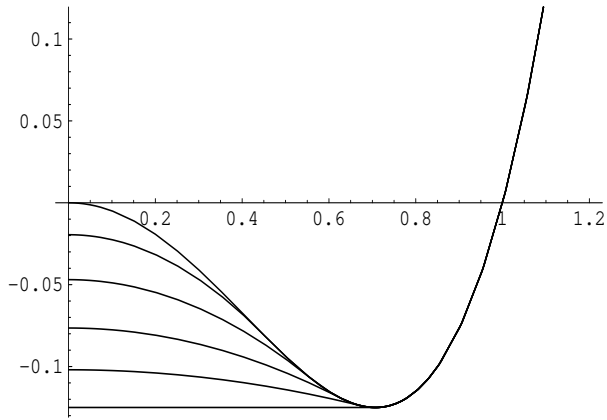}}\hss}
\caption{%
Scaling behaviour of $U_k(A;M^2)$, illustrated by six curves corresponding 
to different values of $k$. With decreasing $k$ the inner region appoaches 
a constant value.} 
\label{f11}
\end{figure}

This singular behaviour
has the effect that the conventional Legendre-transformation is not 
applicable. In such cases one has to refer to the more general 
supremum-definition of the Legendre-transformation, see e.g. \cite{wipfetal}.
In our case it amounts to
\begin{eqnarray}
\label{e17}
U_k(A;M^2)=-k^2\,A^2+\sup\limits_{\varepsilon\ge 0}\left\{2A\,\varepsilon
-F_k(\varepsilon;Q_0)\right\} 
\end{eqnarray}
where
\begin{eqnarray}
\label{e18}
F_k(\varepsilon;Q_0)&\equiv&\frac{1}{{\cal V}}W_k\left[\varepsilon\,
\exp(iQ_{0\mu}
x_\mu+i\beta)\right]\nonumber\\
&=&2\varepsilon\,\tilde{\chi}_0(\varepsilon;Q_0)-\omega_k(M^2)\,
\tilde{\chi}_0^2(\varepsilon;Q_0)-\frac{\lambda}{2}\,\tilde{\chi}_0^4
(\varepsilon;Q_0)
\end{eqnarray}
with $\tilde{\chi}_0$ defined in (\ref{a38,5}).
Using eq. (\ref{a33}) $F_k$ may be rewritten as
\begin{eqnarray}
\label{e19}
F_k(\varepsilon;Q_0)=\frac{3}{2}\varepsilon\,\tilde{\chi}_0(\varepsilon;Q_0)
-\frac{1}{2}\omega_k(M^2)\,\tilde{\chi}_0^2(\varepsilon;Q_0)\,.
\end{eqnarray}
Since $\tilde{\chi}_0(\varepsilon;Q_0)$ is a strictly monotonically increasing
function of $\varepsilon$ we can infer from eq. (\ref{e19}) that $F_k$ is
strictly convex. (This is of course as it should be because in our classical
approximation $W_k$ is related to $S_k$ via a Legendre-transformation.)
Therefore the strict inequality
\begin{eqnarray}
\label{e22}
\frac{\partial F_k}{\partial\varepsilon}(\varepsilon)>
\frac{\partial F_k}{\partial\varepsilon}(\varepsilon=0)\equiv
2\sqrt{-\frac{1}{\lambda}\omega_k(M^2)}\equiv 2\,A_{\rm triv}
\end{eqnarray}
is satisfied for all $\varepsilon>0$. It implies that as long as we have
$A<A_{\rm triv}$ the supremum of $2A\,\varepsilon-F_k(\varepsilon;Q_0)$ is 
always achieved for $\varepsilon=0$ where it has the value $C_k$. As a 
consequence the complete inner region $[0,A_{\rm triv}]$ corresponds to 
$\varepsilon=0$ and the length of this interval is determined by the linear 
term of $F_k$. For $A>A_{\rm triv}$ the
supremum-definition coincides with the familiar definition
of the Legendre-transformation so that eq. (\ref{e17}) leads to
\begin{eqnarray} 
\label{e21}
U_k(A;M^2)=\left\{\begin{array}{ll}C_k-k^2\,A^2
& {\rm if}\;A\le A_{\rm triv}\\ U^{(\rm cl)}(A;M^2) & {\rm if}\;A>A_{\rm triv}
\end{array}\right.\,.
\end{eqnarray}
The behaviour of $U_k$ is illustrated in FIG. \ref{f11} which
contains several curves obtained from eq. (\ref{e21}) for distinct values of
$k$. The case $|p|=M$ is special in that $C_{k=0}$ equals exactly the value
of $U^{(\rm cl)}$ at its minimum $A_{\rm min}^{(\rm cl)}$, and also because it
is precisely $A_{\rm min}^{(\rm cl)}$ which separates the inner from the outer
region, i.e. $A_{\rm triv}=A_{\rm min}^{(\rm cl)}$.\footnote{The potential 
(\ref{e21}) exhibits the special property that $U_k$ 
results from the symmetric vacuum state (\ref{a25}) not only for 
$A<A_{\rm triv}$ but also for $A=A_{\rm triv}$ whereas in the case $|p|\neq M$
the symmetry of the relevant quantum vacuum state is already broken {\it at} 
$A=A_{\rm triv}$. For $|p|=M$ the region of spontaneously broken symmetry is 
restricted to values of $A$ {\it larger} than $A_{\rm triv}$.}
As a consequence the inner region of $U_k$ approaches a constant value
as the cutoff is lowered from $k=k_{\rm cr}$ towards $k=0$. At $k=0$ the inner
region is entirely flat and thus $U_{k=0}$ is found to be convex.
\renewcommand{\baselinestretch}{1}
\small\normalsize
\renewcommand{\baselinestretch}{1.5}
\small\normalsize

Note that, as it should be, the functional $\widetilde{\Gamma}_k$ is always 
convex even in situations where $U_k$, i.e. $\Gamma_k$ is not. Comparing 
(\ref{w1}) to (\ref{d3}) we see that $\widetilde{\Gamma}_k$ yields 
$\widetilde{U}_k\equiv U_k+k^2A^2$ when evaluated for plane waves. For 
instance, from eq. (\ref{e21}) it follows that $\widetilde{U}_k=C_k$ if $A\le 
A_{\rm triv}$ and $\widetilde{U}_k=U^{(\rm cl)}+k^2A^2$ if $A>A_{\rm triv}$, 
which is perfectly convex for any value of $k$.

The physical interpretation of this behaviour of $W_k$ and $\Gamma_k$ is as 
follows. Eq. (\ref{e16}) is nothing but the standard formula
\begin{eqnarray}
\label{x2}
\left.\frac{\delta W_k}{\delta J^*}\right|_{J=0}=\left.\left<\chi\right>
\right|_{J=0}\equiv\left<\chi\right>
\end{eqnarray}
evaluated for plane waves. Let us look at this equation for $k=0$. The
nonvanishing r.h.s. of eq. (\ref{e16}) shows that the modes with $|p|=M$  
acquire a vacuum expectation value. After the source $J$ has been switched off
adiabatically the expectation value
\begin{eqnarray}
\label{x3}
\left<\chi(x)\right>=\frac{M}{\sqrt{2\lambda}}\,\exp\left(iM n_\mu x_\mu
+i\beta\right)
\end{eqnarray}
``remembers'' both the direction $n_\mu$ and the phase $\beta$ of the source.
This singles out a point $(n,\beta)$ of the vacuum manifold and leads to a 
spontaneous breaking of both the {\sf ISO($d$)} symmetry of spacetime rotations
and of the {\sf U(1)} phase symmetry.

This formation of a vacuum condensate happens only for the modes with $|p|=M$
but not for $|p|\neq M$. The difference of the two cases is nicely illustrated
by the plots of the various effective potentials. Let us look at FIG. \ref{f9}
for $|p|\neq M$, say, and let us put a ``ball'' into the minimum of the 
potential $U_k$ at $k=k_{\rm cr}$. Then, when we lower $k$, at a certain point
the ball rolls down from the local minimum at $A\neq 0$ to the minimum at 
$A=0$. Thus, for $k=0$, the corresponding field mode has no expectation value.
The situation is different in FIG. \ref{f11} for $|p|=M$. Until the very last
moment of the evolution the ball always sits at the global minimum of the
potential and has no tendency to roll towards $A=0$. Only for a strictly 
vanishing cutoff $k$, $U_k(A=0;M^2)$ is as low as $U_k$ at the minimum. This
means that the corresponding mode acquires an expectation value. In fact, our
discussion here is remarkably similar to the analogous treatment of the 
familiar spontaneous symmetry breaking by a Mexican-hat potential. In the 
latter case it is the $(p=0)$-modes which condense, but in the language of the 
generalized effective average potentials $U_k$ for arbitrary momenta this
makes no conceptual difference.

Within the present approximation the true vacuum consists of a single plane
wave. Therefore this field configuration serves as a ``master field'' from
which the expectation value of any composite operator can be computed by 
simply inserting the r.h.s. of eq. (\ref{x3}). For $\partial_\mu\chi^*\,
\partial_\mu\chi$, say, this leads to the kinetic condensate (\ref{i2})
announced in the introduction.
\section{Conclusion}
\label{9}
In this paper we investigated a scalar model with a nonstandard inverse 
propagator consisting of a destabilizing $-p^2$-term and a stabilizing 
$+p^4$-term. We find that this model exhibits both spontaneous breaking of
translation symmetry and of a global {\sf U(1)} phase symmetry. The 
ground state
respects, nevertheless, a modified combined translation symmetry which also
involves phase rotations. The rotation symmetry is broken from {\sf SO($d$)} 
to {\sf SO($d-1$)}. In classical or quantum statistical systems our model 
describes
the spontaneous formation of layers in an otherwise homogeneous and isotropic
setting. For such models already a tiny perturbation leads to the formation 
of a geometrical structure.

In order to gain a detailed understanding how the instabilities are removed
from the effective action by including the effects of fluctuations we have
performed a renormalization group analysis. In particular, we have calculated 
the renormalization group flow of the dressed inverse propagator 
$\Sigma_k(p^2)$ for zero fields and of the finite-momentum effective potentials
$U_k(A;p^2)$ in leading order of the semiclassical expansion. We found strong 
renormalization effects which are ``instability induced'' rather than 
``fluctuation induced''. They are driven by the classical instability of the
trivial saddle point in certain regions in the space of field configurations.
This is related to the fact that the global minimum of the Euclidean action
$S[\chi]$ is not at $\chi=0$. Instead, it is realized by nontrivial spin-wave 
configurations which form a space $S^1\times S^{d-1}$ of classical vacua. At 
the level of the
{\it effective} theory, we found that the theory stabilizes itself in a 
dynamical way. The dressed kinetic operator $\Omega_{\rm eff}\equiv
\Sigma_{k=0}$ gives a strictly positive action to all field modes with momenta
$|p|\neq M$. For modes with $|p|=M$ it vanishes. These modes are stabilized by
a ``shift to the true vacuum'' which is similar to what happens in standard
spontaneous symmetry breaking with a Mexican-hat potential. The modes with 
$|p|=M$ form a spatially
nonuniform, Poincar\'{e}- and phase-symmetry breaking condensate. Within the
semiclassical approximation, the true vacuum consists of a single spin wave of 
momentum $M$ and amplitude $M/(2\lambda)$, and is of an obviously 
nonperturbative nature therefore. The fixed phase and direction of this spin
wave lead to a spontaneous breaking of the classical {\sf U(1)}$\times${\sf
ISO($d$)} symmetry.

In this paper we only have considered a model without a classical mass term.
Due to our particular choice of the infrared cutoff ${\cal R}_k=k^2$ 
generalized results for models with a mass term can easily be infered from our
results for nonvanishing $k$.

In the introduction we mentioned that a strong motivation for studying
this model is its similarity with Euclidean quantum gravity based upon actions
such as $\int d^dx\,\sqrt{g}\left\{\alpha\,R+\beta\,R^2\right\}$. For the
conformal factor of the metric such an action contains a negative contribution
to the kinetic energy, coming from $\sqrt{g}R$ and dominating at momenta small
compared to the Planck mass. At large momenta the action becomes positive and
all modes are stable because of the manifestly positive contribution arising
from $\sqrt{g}R^2$. In view of this analogy it is plausible to speculate that
also quantum gravity dynamically stabilizes the conformal factor by developing
a nontrivial vacuum structure, with nonzero condensates such as $\left<
(\bar{D}_\mu\phi)^2\right>$, so that all excitations about this ground state
are stable.

In the scalar model the semiclassical expansion about $\chi_{\rm min}$, the
global minimum of $S_k^J$, has led to an effective kinetic operator 
$\Omega_{\rm eff}(p^2)$ which has stabilized (almost) all modes which were 
unstable with 
respect to the classical $\Omega(p^2)$. In gravity we might expect a similar
mechanism to be at work when we expand about the global minimum of $\int d^dx
\,\sqrt{g}\left\{\alpha\,R+\beta\,R^2\right\}$. Roughly speaking, leaving finer
details of the momentum dependence aside, the $\Omega_{\rm eff}(p^2)$-curve is
obtained from $\Omega(p^2)$ by shifting it upward by a constant 
$(\rm mass)^2$-term $M^2/2$, see FIG. \ref{f13}. So the dynamical stabilization
of the scalar model is essentially a ``mass generation''.

This mass generation also provides the justification for our loop expansion 
and re\-tai\-ning the lowest order contribution only. Contrary to the case
of massless models with an ordinary kinetic term where the loop expansion 
does not lead to reliable results \cite{CW}, 
in our model the mass generation cuts off loops so that the loop correction to 
$\lambda$, for instance, is negligible.

It is an important
question how a similar mass generation would manifest itself in the effective
average action for gravity, $\Gamma_k[g_{\mu\nu}]$, and which type of 
truncations should be used in order to obtain it from the flow equation 
\cite{aagrav}. It is clear that a naive mass term for the conformal factor is
forbidden by general coordinate invariance. But also local curvature invariants
$R^2$, $R_{\mu\nu}R^{\mu\nu}$ etc. are of no help because they vanish for flat
space and will not lead to an effective action whose minimum is at $g_{\mu\nu}
=\delta_{\mu\nu}$ \cite{nonloc}. This suggests that the relevant terms in 
$\Gamma$ and $\Gamma_k$ must be {\it nonlocal} if expressed in a gauge 
invariant way. (After gauge fixing, they may be local, nevertheless.) For 
instance, a higher 
dimensional analogue of the $2d$ induced gravity action $\int d^dx\,\sqrt{g}\,
R\,\Box^{-1}\,R$, added to the Einstein-Hilbert term, is known to have flat 
space as its global minimum \cite{nonloc}. Hence all fluctuations about this
ground state, including those of the conformal factor, are stable. Therefore
it would be very interesting to study the renormalization group flow of
$\Gamma_k[g_{\mu\nu}]$ using a truncation of the space of actions which
includes nonlocal invariants. Work along these lines is in progress.

One may wonder if the analogy between our model and gravity can be put even
further. In a gauge fixed version of gravity the local symmetry of general
coordinate transformations may be ``spontaneously broken'', similar to the
Higgs-picture for local gauge theories. This language is usually employed in
order to describe spontaneous compactification of higher dimensional theories.
The fact that the minimum of the Euclidean action occurs for a 
non-translationally invariant field configuration then strongly suggests the
existence of additional space dimensions, $d>4$. Otherwise, for $d=4$, the
spectrum of excitations may not exhibit the full four-dimensional 
Poincar\'{e} symmetry, similar to the spectrum shown in FIG. \ref{f7}. In
higher dimensions, the $d$-dimensional Poincar\'{e} symmetry may be reduced to
a four-dimensional Poincar\'{e} symmetry, again similar to our example. 
Actually, classical solutions with spontaneous compactification which have a
lower Euclidean action than flat $d$-dimensional space have been discussed a
long time ago \cite{WRW}. In view of the present paper it would be very 
interesting to find realistic classical solutions corresponding to the
absolute minimum of the Euclidean action.
\begin{appendix}
\section{Global minimum for plane-wave sources}
\label{Acomplete}
In this part of the appendix we concentrate on determining the global minimum
of the action $S_k^J$ for plane-wave sources $J=\varepsilon\,
\exp\left(i p_\mu x_\mu+i
\beta\right)$. In the first subsection we discuss two kinds of
solutions, each of them yielding the global minimum in a certain range of
the $(\varepsilon,p,k)$-parameter space. In the second subsection
the function $\varepsilon_k(p^2)$, which describes the region in the 
parameter space separating these ranges, is exactly derived. The third 
subsection of this appendix contains additional calculational details needed 
for obtaining some of the results given in the first subsection.
\subsection{Solutions of the e.o.m.}
\label{A}
For the calculation of the effective average action for plane-wave average 
fields (see section \ref{8}) it is necessary to find out some properties of 
the solutions corresponding to nonvanishing sources which are plane waves of 
the form
\begin{eqnarray}
\label{29}
J(x)=\varepsilon\,\exp\left(i p_\mu x_\mu+i\beta\right)
\end{eqnarray}
with a real ``amplitude'' $\varepsilon\ge 0$ and $\beta\in [0,2\pi)$.
This restriction allows us to calculate the minimizing field configurations 
either exactly (for $k^2\ge M^2/2$) or at least approximately for small values
of $\varepsilon$ (for $k^2\le M^2/2$).

For the source (\ref{29}), the e.o.m. we have to solve takes the form
\begin{eqnarray}
\label{a30}
\left[\Omega(-\Box)+k^2+\lambda\,\left|\chi\right|^2\right]\chi=\varepsilon
\,\exp\left(i p_\mu x_\mu+i\beta\right).
\end{eqnarray}
\subsubsection{The solution $\chi\propto\exp(ipx+i\beta)$}
The simplest solution one can think of is a field $\chi$ which does not 
``know'' about the existence of the nontrivial, degenerate minimum found for 
$J=0$ and oscillates with the same frequency and phase as the source. If we 
insert the corresponding ansatz
\begin{eqnarray}
\label{a31}
\chi(x)=\chi_0\,\exp\left(i p_\mu x_\mu+i\beta\right)
\end{eqnarray}
into the e.o.m., the result is the $x$-independent equation
\begin{eqnarray}
\label{a32}
\omega_k(p^2)\,\chi_0+\lambda\,\left|\chi_0\right|^2\,\chi_0=\varepsilon\,.
\end{eqnarray}
As we chose $\varepsilon$ to be real, $\chi_0$ must also be real, so that eq. 
(\ref{a32}) boils down to a simple cubic equation in the real variable 
$\chi_0$:
\begin{eqnarray}
\label{a33}
\chi_0^3+\frac{\omega_k(p^2)}{\lambda}\,\chi_0-\frac{\varepsilon}{\lambda}=0
\end{eqnarray}
The general solution of this equation is given by
\begin{eqnarray}
\label{a34}
\chi_0(\varepsilon;p)=\left(\frac{\varepsilon}{2\lambda}
+\sqrt{\frac{\varepsilon^2}{4\lambda^2}
+\frac{\omega_k^3(p^2)}{27\lambda^3}}\right)^{1/3}
+\left(\frac{\varepsilon}{2\lambda}-\sqrt{\frac{\varepsilon^2}{4\lambda^2}
+\frac{\omega_k^3(p^2)}{27\lambda^3}}\right)^{1/3}\,.
\end{eqnarray}
We have to distinguish the two cases where 
\begin{eqnarray}
\label{a118}
\Delta\equiv\frac{\varepsilon^2}{4\lambda^2}+\frac{\omega_k^3(p^2)}
{27\lambda^3}
\end{eqnarray}
is either positive or negative. If it is positive, the above amplitude
$\chi_0(\varepsilon;p)$ represents a single real solution of eq. (\ref{a33}). 
But if it is negative, the square root of $\Delta$ becomes imaginary, 
so that $\chi_0(\varepsilon;p)$ comprises three different real solutions. 
Those can be rewritten in the manifestly real form
\begin{eqnarray}
\label{a36}
\chi_0^{(n)}(\varepsilon;p)=2\sqrt{-\frac{\omega_k(p^2)}{3\lambda}}\,\cos\left[
\frac{1}{3}\arccos\left(\frac{\varepsilon}{2}\sqrt{-\frac{27\lambda}
{\omega_k^3(p^2)}}\right)+\frac{2\pi n}{3}\right]\;,\;\;n=0,1,2.
\end{eqnarray}
However, one can check easily that the only candidate for the global minimum 
is $\chi_0^{(n=0)}$, because the action corresponding to this branch is lower 
than the action corresponding to the other two branches.\footnote{The 
appertaining proof can be found in appendix \ref{B}.} Taking into
account that the branch $\chi_0^{(n=0)}$ is the only one that coincides with 
eq. (\ref{a34}) at $\Delta=0$, this result is not very surprising.

Combining the above expressions, which describe potential minima in the two
complementary regions of $\Delta$, we can formulate solutions for the whole 
range of $\Delta$ and thus also of $\varepsilon$. We have to consider two 
distinct cases. For $\omega_k(p^2)\ge 0$, $\Delta$ is always nonnegative, so 
the candidate for the global minimum reads 
\begin{eqnarray}
\label{a37}
\chi_{\rm min}(x)=\left[\left(\frac{\varepsilon}{2\lambda}
+\sqrt{\Delta}\right)^{1/3}
+\left(\frac{\varepsilon}{2\lambda}-\sqrt{\Delta}\right)^{1/3}\right]\,
\exp\left(i p_\mu x_\mu+i\beta\right).
\end{eqnarray}
For $\omega_k(p^2)<0$, we have to fit together the two relevant solutions for 
$\Delta\ge 0$ and $\Delta<0$, so that
\begin{eqnarray}
\label{a38}
\chi_{\rm min}(x)=\tilde{\chi}_0(\varepsilon;p)\,\exp\left(i p_\mu x_\mu
+i\beta\right)
\end{eqnarray}
where
\begin{eqnarray}
\label{a38,5}
\tilde{\chi}_0(\varepsilon;p)=\left\{\begin{array}{c}
\left(\frac{\varepsilon}{2\lambda}+\sqrt{\Delta}\right)^{1/3}
+\left(\frac{\varepsilon}{2\lambda}-\sqrt{\Delta}\right)^{1/3}\;,\;\;
\Delta\ge 0\\
2\sqrt{-\frac{\omega_k(p^2)}{3\lambda}}\,\cos\left[
\frac{1}{3}\arccos\left(\frac{\varepsilon}{2}\sqrt{-\frac{27\lambda}
{\omega_k^3(p^2)}}\right)\right]
\;,\;\;\Delta<0\end{array}\right.\,.
\end{eqnarray}
\subsubsection{Is $\chi\propto\exp(ipx+i\beta)$ the global minimum?}
What remains to be dealt with is the question whether the solutions 
(\ref{a37}), (\ref{a38}) constitute absolute minima or just saddle points. 
First we check this for large values of $k$, $k^2\ge M^2/2$:\\[12pt]
In this region the relation $\omega_k(p^2)\ge 0$ is always satisfied which
means that $\Delta$ cannot become negative. If we take the corresponding 
solution, eq. (\ref{a37}), and send $\varepsilon$ to zero, we end up with 
$\chi_{\rm min}=0$, which was already shown to be the minimizing field
configuration for $J=0$. In addition, it is clear that the integral
\begin{eqnarray}
\label{a39}
\int\frac{d^dq}{(2\pi)^d}\,\left|\widetilde{\delta\chi}(q)\right|^2
\,\left\{\Omega(q^2)+k^2+\lambda\,
\left[\left(\frac{\varepsilon}{2\lambda}+\sqrt{\Delta}\right)^{1/3}
+\left(\frac{\varepsilon}{2\lambda}-\sqrt{\Delta}\right)^{1/3}\right]^2\right\}
\end{eqnarray}
is always nonnegative (remember that $\omega_k(q^2)=\Omega(q^2)+k^2\ge 0$ for 
$k^2\ge M^2/2$) and thus $\Delta S_k\ge 0$ is satisfied.\footnote{The 
momentum $q$ is not to be confused with the momentum of the source $p$.} 
Consequently we can identify (\ref{a37}) as the field that corresponds to the 
absolute minimum for $k^2\ge M^2/2$; it coincides with the result of section
\ref{4} if we set $\varepsilon=0$.

Next we investigate the case $k^2< M^2/2$. 
The amplitudes of the solutions corresponding to both $\omega_k(p^2)\ge 0$ and 
$\omega_k(p^2)<0$ grow monotonically, as we increase $\varepsilon$, without 
approaching any finite bound. This means
that, for sufficiently large values of $\varepsilon$, 
$\lambda|\chi_{\rm min}|^2$ will always compensate any possible negative
value of $\omega_k(q^2)\ge -M^2/2$, rendering the integrals (\ref{a39})
or  
\begin{eqnarray}
\label{a40}
\int\frac{d^dq}{(2\pi)^d}\,\left|\widetilde{\delta\chi}(q)\right|^2
\,\left[\Omega(q^2)+k^2+\lambda\,\tilde{\chi}_0^2(\varepsilon;p)\right]
\end{eqnarray}
and thus $\Delta S_k$ positive. On the other hand the solutions 
(\ref{a37}), (\ref{a38}) do not approach the nontrivial, degenerate solution 
(\ref{a25}) as we send $\varepsilon$ to zero. (Even for momenta 
$p_\mu=M n_\mu$, $|n|=1$, the limit $\varepsilon\rightarrow 0$ produces 
only a unique solution of the form (\ref{a25}), 
with the unit vector $n$ and the phase $\beta$ fixed.) In view of this 
behaviour we can state that for any momentum $p$ with $|p|\neq M$ there exists
a certain value $\varepsilon_k(p^2)>0$, so that for all $\varepsilon\ge 
\varepsilon_k(p^2)$ one of the solutions (\ref{a37}) or (\ref{a38}) yields the 
absolute minimum; it depends on the value of $\omega_k(p^2)$ whether 
(\ref{a37}) or (\ref{a38}) is the right one. This implies that for 
$\varepsilon$-values below the corresponding boundary value $\varepsilon_k(p^2
)$ any of the above solutions becomes unstable under certain deformations and
therefore is a saddle point or a local extremum. For sufficiently small
 $\varepsilon$-values
the true global minimum must be a generalization of the nontrivial minimum we 
calculated for the case of vanishing sources.

{\bf An upper bound for $\varepsilon_k$}: 
Due to the condition (\ref{a13}) we can find an upper bound 
$\tilde{\varepsilon}_k(p^2)$ for the amplitude $\varepsilon_k(p^2)$, where the 
transition from one $\varepsilon$-region to the other takes place. All we have
to do is to insert $\chi_0=\hat{\chi}_0$ - this is the lowest value for 
$\chi_0$, for which the first integral of eq. (\ref{a15}) is manifestly 
positive (or zero) - into
eq. (\ref{a33}) and calculate the corresponding amplitude $\varepsilon$. The 
result is 
\begin{eqnarray}
\label{a41}
\tilde{\varepsilon}_k(p^2)=\frac{1}{2M^2}\,\sqrt{-
\omega_k(M^2)/\lambda}\left(p^2-M^2\right)^2.
\end{eqnarray}
We can turn the above condition on $\varepsilon$ into
a condition on the momentum of the source $p$. It follows immediately that for
all $\varepsilon$ there exists a momentum $p_k(\varepsilon)$, so that for
all $p^2$ satisfying the relation
\begin{eqnarray}
\label{a51}
\max(0,M^2-p_k^2(\varepsilon))\le p^2\le M^2+p_k^2(\varepsilon)
\end{eqnarray}
the solutions (\ref{a37}) and (\ref{a38}) represent absolute minima. A lower
bound $\tilde{p}_k^2(\varepsilon)$ for $p_k^2(\varepsilon)$ is given by the 
expression
\begin{eqnarray}
\label{a52}
\tilde{p}_k^2(\varepsilon)=\sqrt{\frac{2M^2\,\varepsilon}{\sqrt{-
\omega_k(M^2)/\lambda}}}\le p_k^2(\varepsilon).
\end{eqnarray}
(In appendix \ref{A1} we proof that indeed $\tilde{\varepsilon}_k(p^2)=
\varepsilon_k(p^2)$ and thus $\tilde{p}_k^2(\varepsilon)=p_k^2(\varepsilon)$.)

{\bf The exceptional case $|p|=M$}: For momenta $p_\mu=Q_{0\mu}=M n_\mu$, 
$|n|=1$, the situation is  
more subtle. For all $\varepsilon>0$ we have $\tilde{\chi}_0(\varepsilon;Q_0)
>\hat{\chi}_0$. Hence, if $\varepsilon>0$, the corresponding solution 
(\ref{a38}) represents the absolute minimum. The phase and the direction
of the momentum vector of this solution are uniquely determined by the
corresponding parameters of the source. This is still the case for 
$\varepsilon\rightarrow 0$, where the solution takes the form $\hat{\chi}_0\,
\exp(iM n_\mu x_\mu+i\beta)$ with $n_\mu$, $\beta$ fixed. Despite of the fact
that the integral (\ref{a40}) is nonnegative for this solution, it does not  
really yield the true vacuum for vanishing sources, which is degenerate 
with respect to the phase and the direction of the momentum $Q_0$. Thus we 
recognize a discontinous behaviour concerning the degeneracy of
the true vacuum when we switch on sources of the form $\varepsilon
\,\exp(i Q_{0\mu} x_\mu+i\beta)$. This is analogous to the ``tilting'' of a
Mexican-hat potential caused by a symmetry breaking source.
\subsubsection{Perturbative expansion about $\chi_{\rm min}(J=0)$}
Let us now return to the case $p_\mu\neq M n_\mu$. As mentioned before, the 
solution for sufficiently small $\varepsilon$ has to depend on 
$\varepsilon$ in such a way that it approaches the degenerate solution 
(\ref{a25}) for $\varepsilon\rightarrow 0$. As we show below it is a function 
of the same free parameters as (\ref{a25}) and is degenerate as well, with the 
vacuum manifold given by $S^1\times S^{d-1}$. 

There are two important points concerning this solution that we do not 
really know. Neither do we know if the transition ``degenerate 
$\leftrightarrow$ nondegenerate'' is discontinuous as in the case above 
(a smoothly vanishing dependence on the free parameters would be conceivable 
as well), nor do we know the point where this transition takes place, since we
have no proof
that there are no intermediate solutions connecting the degenerate solution to
the solution which is valid for $\varepsilon\ge\varepsilon_k(p^2)$. It is 
natural to assume that no such intermediate solutions exist and that the 
transition point is given by $\varepsilon_k(p^2)$. This would mean that, 
contrary to the case $p_\mu =M n_\mu$, the degeneration occurs {\it below} and 
not {\it at} a certain  boundary value of $\varepsilon$, since the source is 
nonzero at $\varepsilon_k(p^2)$ and still dictates the phase and the direction
of the momentum. 

From now on we will identify the domain of validity
of this solution for small values of $\varepsilon$ with the region
$\varepsilon<\varepsilon_k(p^2)$. One should bear in mind that in principle 
there could be additional, intermediate solutions in this region. However, the
numerical evidence which we present in section \ref{8} strongly supports the
assumption that those intermediate solutions do not exist.

Let us have a closer look at the structure of the solution for $\varepsilon<
\varepsilon_k(p^2)$. Due to the information we have about this solution we
may expand it in a power series of the form
\begin{eqnarray}
\label{a42}
\chi_{\rm min}(x)=\hat{\chi}_0\,\exp\left(i Q_{0\mu} x_\mu+i\alpha\right)
+\sum\limits_{n=1}^{\infty}\varepsilon^n\,\varphi^{(n)}(x)\,.
\end{eqnarray}
In order to calculate the first order term $\varphi^{(1)}$, we insert the 
ansatz
\begin{eqnarray}
\label{a43}
\chi_{\rm min}(x)=\hat{\chi}_0\,\exp\left(i Q_{0\mu} x_\mu+i\alpha\right)
+\varepsilon\,\varphi^{(1)}(x)+{\cal O}(\varepsilon^2)
\end{eqnarray}
into the e.o.m. and obtain at order $\varepsilon$
\begin{eqnarray}
\label{a44}
& &\left[\omega_k(-D^2)-2\omega_k(M^2)\right]\,\varphi^{(1)}\,
\exp\left(-i Q_{0\mu}x_\mu-i\alpha\right)-\omega_k(M^2)\,\varphi^{(1)*}\,
\exp\left(i Q_{0\mu}x_\mu+i\alpha\right)\nonumber\\
& &=\exp\left(i (p_\mu-Q_{0\mu})x_\mu+i(\beta-\alpha)\right)
\end{eqnarray}
where
\begin{eqnarray}
\label{101}
D_\mu\equiv\partial_\mu+i Q_{0\mu}\,.
\end{eqnarray}
It is convenient to introduce the field
\begin{eqnarray}
\label{a45}
\psi(x)\equiv\varphi^{(1)}(x)\,\exp\left(-i Q_{0\mu}x_\mu-i\alpha\right)
\end{eqnarray}
in terms of which equation (\ref{a44}) looks more transparent:
\begin{eqnarray}
\label{a46}
\left[\omega_k(-D^2)-2\omega_k(M^2)\right]\,\psi
-\omega_k(M^2)\,\psi^*=\exp\left(i (p_\mu-Q_{0\mu})x_\mu+i(\beta-\alpha)\right)
\end{eqnarray}
Obviously the most general solution to this equation can be obtained from the
ansatz
\begin{eqnarray}
\label{a47}
\psi(x)=f\,\exp\left(i (p_\mu-Q_{0\mu})x_\mu+i(\beta-\alpha)\right)
+g\,\exp\left(-i (p_\mu-Q_{0\mu})x_\mu-i(\beta-\alpha)\right)\,.
\end{eqnarray}
After some simple manipulations we find the following expressions for the 
parameters $f$ and $g$:
\begin{eqnarray}
\label{a48}
f=\frac{\omega_k\left((2Q_0-p)^2\right)-2\omega_k(M^2)}
{\left[\omega_k\left((2Q_0-p)^2\right)
-2\omega_k(M^2)\right]\left[\omega_k(p^2)-2\omega_k(M^2)\right]
-\omega_k^2(M^2)}
\end{eqnarray}
\begin{eqnarray}
\label{a49}
g=\frac{\omega_k(M^2)}{\left[\omega_k\left((2Q_0-p)^2\right)
-2\omega_k(M^2)\right]\left[\omega_k(p^2)-2\omega_k(M^2)\right]
-\omega_k^2(M^2)}
\end{eqnarray}
By inserting the above expressions into eq. (\ref{a47}) and 
multiplying the result by $\exp(i Q_{0\mu}x_\mu+i\alpha)$ we obtain the desired
expression for $\varphi^{(1)}$. Thus, in the region $k^2\le M^2/2$, the 
absolute minimum of the action $S_k^J$, containing sources with 
sufficiently small amplitudes $\varepsilon$, reads
\begin{eqnarray}
\label{a50}
\lefteqn{\chi_{\rm min}(x)=\hat{\chi}_0\,\exp\left(i Q_{0\mu}x_\mu+i\alpha
\right)}\nonumber\\
& &+\varepsilon\,
\frac{\left[\omega_k\left((2Q_0-p)^2\right)-2\omega_k(M^2)\right]\,
+\omega_k(M^2)
\,\exp\left(2i (Q_{0\mu}-p_\mu)x_\mu+2i(\alpha-\beta)\right)}
{\left[\omega_k\left((2Q_0-p)^2\right)
-2\omega_k(M^2)\right]\left[\omega_k(p^2)-2\omega_k(M^2)\right]
-\omega_k^2(M^2)}\nonumber\\
& &\times\exp\left(ip_\mu x_\mu+i\beta\right)+{\cal O}(\varepsilon^2)
\end{eqnarray}
which is of course equivalent to eq. (\ref{a61}) combined with (\ref{a69}).
Like the free one, the minimum (\ref{a50}) is parametrized by the directions 
of $Q_0$ and the phase $\alpha$. At the ``resonance'' $|p|=M$ the expansion
(\ref{a50}) is not well defined for all directions of $p_\mu$ which, again,
illustrates why this case is special; see the discussion following eq. 
(\ref{a52}).
\subsection{A necessary condition for the absolute minimum}
\label{A1}
A given field configuration $\chi_{\rm min}$ minimizes the action $S_k^J$
globally if and only if $\Delta S_k[\chi_{\rm min},\delta\chi]$, defined by
eq. (\ref{a12}), is nonnegative for all deformations $\delta\chi$. In section
\ref{4} we decomposed $\Delta S_k$ appropriately and derived a sufficient
condition, eq. (\ref{a100}), which tells us that $\Delta S_k$ is always
nonnegative for plane waves $\chi_{\rm min}=\chi_0\,\exp(i p_\mu x_\mu
+i\alpha)$ provided that $k^2\ge M^2/2$ or $\chi_0\ge \hat{\chi}_0$ if
$k^2<M^2/2$. In this section we proof that, {\it for plane-wave sources},
this condition is also {\it necessary}, i.e. that, for $k^2<M^2/2$, the
solutions (\ref{a37}) and (\ref{a38}) represent saddle points rather than
absolute minima if $\chi_0(\varepsilon;p)<\hat{\chi}_0$,
$\tilde{\chi}_0(\varepsilon;p)<\hat{\chi}_0$. We show that there
always exist certain (infinitesimal) deformations which render $\Delta S_k$
negative.

We start our proof by writing down $\Delta S_k$ in the form
\begin{eqnarray}
\label{c0}
\Delta S_k\left[\chi_{\rm min},\delta\chi\right]&=&\frac{1}{2}\int d^dx\,
\left(\delta\chi^*,\delta\chi\right)\,\hat{S}_k^{(2)}\left[\chi_{\rm min},
\delta\chi
\right]\,\left(\begin{array}{c}\delta\chi\\\delta\chi^*\end{array}\right)
\nonumber\\
& &+\frac{\lambda}{2}\int d^dx\,\left\{2\left(\chi_{\rm min}^*\,\delta\chi
+\chi_{\rm min}\,\delta\chi^*\right)|\delta\chi|^2+|\delta\chi|^4\right\}.
\end{eqnarray}
After inserting $\chi_{\rm min}=\chi_0\,\exp(ip_\mu x_\mu+i\beta)$
into (\ref{c0}) we diagonalize the matrix operator $\hat{S}_k^{(2)}$ and obtain
in analogy with appendix \ref{D} for the part of $\Delta S_k$ which is
quadratic in the deformations
\begin{eqnarray}
\label{c1}
\lefteqn{\Delta S_k^{\rm quad}\left[\chi_0\,\exp(ip_\mu x_\mu+i\beta),
\delta\chi\right]}\nonumber\\
&=&\frac{1}{2}\int d^dx
\left\{\Psi_1(\delta\chi)\,\tilde{\Lambda}_{k,1}(-D_p^2)\,\Psi_1(\delta\chi)
+\Psi_2(\delta\chi)\,\tilde{\Lambda}_{k,2}(-D_p^2)\,\Psi_2(\delta\chi)\right\}
\end{eqnarray}
where
\begin{eqnarray}
\label{c2}
\tilde{\Lambda}_{k,1/2}(-D_p^2)&\equiv &\omega_k(D_p^2)+\omega_k(D_p^{*2})+4
\lambda\,\chi_0^2\mp\sqrt{4\lambda^2\,\chi_0^4+\left(\Omega(-D_p^{*2})
-\Omega(-D_p^2)\right)^2}\nonumber\\
D_{p\mu}&\equiv &\partial_\mu+ip_\mu\,;\;\;D_{p\mu}^*\equiv\partial_\mu-i
p_\mu\,.
\end{eqnarray}
The {\it real} fields $\Psi_1$ and $\Psi_2$ depend on $\delta\chi$ by
relations similar to those between $K_{1/2}$ and $J$ given by eq.
(\ref{-a25}). They may be treated as new, independent variables.

It is important to note that the operator $\tilde{\Lambda}_{k,1}(-D_p^2)$,
when applied to $\exp(\pm i q_\mu x_\mu)$ with $q_\mu$ perpendicular to
$p_\mu$ $(q_\mu p_\mu=0)$, yields the expression
\begin{eqnarray}
\label{c3}
\tilde{\Lambda}_{k,1}((q_\mu\pm p_\mu)^2)=-2(q^2+p^2)+\frac{1}{M^2}\left(q^2+
p^2\right)^2+2k^2+2\lambda\,\chi_0^2\,.
\end{eqnarray}
For $q^2=M^2-p^2$ we obtain 
\begin{eqnarray}
\label{c4}
\tilde{\Lambda}_{k,1}((q_\mu\pm p_\mu)^2)=2k^2-M^2+2\lambda\,\chi_0^2
\end{eqnarray}
which is obviously negative for all $\chi_0<\hat{\chi}_0 \equiv
\sqrt{(M^2-2k^2)/(2\lambda)}$. 
This implies that (for all $\chi_0<\hat{\chi}_0$) $\Delta S_k^{\rm quad}<0$
can be achieved by any deformation of the form
\begin{eqnarray}
\label{c5}
(\Psi_1,\Psi_2)=A\left\{\begin{array}{ll}
\left(\cos(\sqrt{M^2-p^2}\, m_\mu x_\mu),0\right) & {\rm if}\;p^2<M^2\\
\left(\exp(-\sqrt{p^2-M^2}\,m_\mu x_\mu),0\right) & {\rm if}\;p^2>M^2 
\end{array}\right.
\end{eqnarray}
where $A$ represents a nonvanishing, real parameter and $m_\mu$ is a unit
vector perpendicular to $p_\mu$.
It is not difficult to show that the above deformations (\ref{c5}) are
related to our original deformations $\delta\chi$ via
\begin{eqnarray}
\label{c6}
\delta\chi(x)=\Psi_1(x)\,\exp(ip_\mu x_\mu+i\beta)\,.
\end{eqnarray}
Thus inserting eqs. (\ref{c5}), (\ref{c6}) into $\Delta S_k$ leads to
\begin{eqnarray}
\label{c7}
\lefteqn{\Delta S_k\left[\chi_0\,\exp(ip_\mu x_\mu+i\beta),
\Psi_1(x)\,\exp(ip_\mu x_\mu+i\beta)\right]}\nonumber\\
&=&\frac{1}{2}\int d^dx\left\{\left(2k^2-M^2+2\lambda\,\chi_0^2\right)\,
\Psi_1^2(x)+4\lambda\,\chi_0\,\Psi_1^3(x)+\lambda\,\Psi_1^4(x)\right\}
\end{eqnarray}
By putting the system in a box with a finite volume ${\cal V}\equiv\int d^dx$,
we may now introduce an appropriate ${\cal V}$-dependent amplitude 
$\delta/{\cal N}({\cal V})$ instead of the parameter $A$ such that, in the 
limit ${\cal V}\rightarrow\infty$,
the integral $\int d^dx\,\Psi^n_1(x)$ remains finite for $n=2,3,4$ and
is $\neq 0$ for $n=2$. This means that for all sufficiently small values of
the (${\cal V}$-independent) parameter $\delta>0$ the terms of third and 
fourth order in $\Psi_1$ are negligible, which leads to $\Delta S_k<0$ if 
$\chi_0<\hat{\chi}_0$.

A possible choice for ${\cal N}({\cal V})$ is given by
\begin{eqnarray}
\label{c8}
{\cal N}({\cal V})=\left\{\begin{array}{ll}\sqrt{{\cal V}} & {\rm if}\;
p^2<M^2\\ C({\cal V}_\nu)
\int d^dx\,\exp(-\sqrt{p^2-M^2}\,m_\mu x_\mu) & {\rm if}\;p^2>M^2\end{array}
\right.
\end{eqnarray}
where ${\cal V}_\nu \equiv\int dx_\nu$ and $C({\cal V}_\mu)=1$ if 
$m_\nu\neq 0$ for all 
$\nu$, $C({\cal V}_\mu)=1/(\prod\limits_{\nu:m_\nu=0}\sqrt{{\cal V}_\nu})$ 
otherwise. 

From the above result we deduce that, in case of $k<k_{\rm cr}$, $\hat{\chi}_0$
equals the
boundary value for $\chi_0(\varepsilon;p)$ and $\tilde{\chi}_0(\varepsilon;p)$
which separates the region where the plane wave solutions (\ref{a37}) and
(\ref{a38}) represent the absolute minimum from the region where the
absolute minimum is given by the degenerate solution (\ref{a42}).
This implies that $\tilde{\varepsilon}_k(p^2)=\varepsilon_k(p^2)$.

\subsection{How to choose the correct $\chi_0^{(n)}$}
\label{B}
In this part of the appendix we present the still missing prove that only one
of the three solutions 
\begin{eqnarray}
\label{a200}
\chi_0^{(n)}(\varepsilon;p)\,\exp\left(ip_\mu x_\mu+i\beta\right)\;,\;\;
n=0,1,2\;,
\end{eqnarray}
that we found in the region, where $\omega_k(p^2)<0$ is satisfied,
constitutes a genuine candidate for the global minimum of $S_k^J$, and that is
the solution for $n=0$. What we have to do here is to show that this solution
produces a lower action than the other two. We start the proof by recalling  
the expression for the amplitude $\chi_0^{(n)}(\varepsilon;p)$, which takes the
form
\begin{eqnarray}
\label{a119}
\chi_0^{(n)}(\varepsilon;p)=2\sqrt{-\frac{\omega_k(p^2)}{3\lambda}}\,\cos\left(
\frac{\theta+2\pi n}{3}\right)\;,\;\;n=0,1,2\;,
\end{eqnarray}
where $\theta$ is defined as
\begin{eqnarray}
\label{a120}
\theta=\arccos\left(\frac{\varepsilon}{2}\,\sqrt{-\frac{27\,\lambda}
{\omega_k^3(p^2)}}\right).
\end{eqnarray}
Inserting (\ref{a200}) into $S_k^J$ leads to  
\begin{eqnarray}
\label{a125}
S_k^J\left[\chi_0^{(n)}(\varepsilon;p)\,\exp(ip_\mu x_\mu+i\beta);J\right]
={\cal V}\,\frac{4\omega_k^2(p^2)}{3\,\lambda}\,\left[-2z^4+z^2\right].
\end{eqnarray}
where we introduced the parameter
\begin{eqnarray}
\label{a122}
z\equiv\cos\left(\frac{\theta+2\pi n}{3}\right),
\end{eqnarray}
and used the relation
\begin{eqnarray}
\label{a124}
\cos\theta=4z^3-3z\;.
\end{eqnarray}
The function $f(z)=-2z^4+z^2$
has zeros at $z=-1/\sqrt{2}$, $z=0$ and $z=1/\sqrt{2}$ and exhibits a local
minimum at $z=0$ and two absolute maxima at $z=\pm 1/2$.
This means that $f(z)$ describes a reverse double well. If we
know the range of values, which $z$ covers for $n=1,2,3$, we can use this
information concerning the behaviour of $f(z)$ to show that the solution for
$n=0$ always yields the least action. For the $z$-intervals we need to know 
we find:
\begin{eqnarray}
\label{a127}
\begin{array}{lll}
n=0\;: & \arccos z\in\left[0,\frac{\pi}{6}\right]\;\Rightarrow & z\in\left[
\frac{\sqrt{3}}{2},1\right]\\
n=1\;: & \arccos z\in\left[\frac{2\pi}{3},\frac{5\pi}{6}\right]\;\Rightarrow &
z\in\left[-\frac{\sqrt{3}}{2},-\frac{1}{2}\right]\\
n=3\;: & \arccos z\in\left[\frac{4\pi}{3},\frac{3\pi}{2}\right]\;\Rightarrow &
z\in\left[\frac{-1}{2},0\right]
\end{array}
\end{eqnarray} 
Since the symmetric function $f(z)$ grows monotonically for 
$z\in(-\infty,-1/2]$, decreases monotonically for $z\in [1/2,\infty)$ 
and is smaller than zero only for 
$z\in(-\infty,-1/\sqrt{2})\cup (1/\sqrt{2},\infty)$ one realizes immediately 
that the solution corresponding to $n=0$ always produces the least action,
except for the $z$-values belonging to the angle $\theta=\pi/2$, where
the action for $n=0$ is identical with the one for $n=1$. Thus we have 
proved that our statement is correct.
\section{Symmetry breaking by a fixed spin-wave configuration}
\label{D}
In this appendix we diagonalize the matrix differential operator
$\hat{S}^{(2)}_k[\varphi^{(0)}]^{-1}$ for $\varphi^{(0)}\equiv
\chi_{\rm min}(J=0)$ given by (\ref{a25}). From the technical point of view
this amounts to the computation of $W_k$ or $\Gamma_k$ with the integration
over the vacuum manifold omitted, i.e. we consider a plane wave
$\varphi^{(0)}$ with a fixed direction $n_\mu$ and phase $\alpha$.

We start from the definition
\begin{eqnarray}
\label{b4}
W_k^{n,\alpha}\left[J\right]\equiv\frac{1}{2}\int d^dx \left(J^*,J\right)\,
\hat{S}^{(2)}_k\left[\varphi^{(0)}\right]^{-1}\,\left(\begin{array}{c}
J\\ J^*\end{array}\right)+{\cal O}(J^2J^{*2})
\end{eqnarray}
which is analogous to (\ref{-a17}) but does not include an integration over
$n_\mu$ and $\alpha$. Then, diagonalizing the operator 
$\hat{S}_k^{(2)}[\varphi^{(0)}]^{-1}$ via the unitary transformation
\begin{eqnarray}
\label{-a23}
\hat{S}_k^{(2)}[\varphi^{(0)}]^{-1}&\rightarrow & V U\, 
\hat{S}_k^{(2)}[\varphi^{(0)}]^{-1}\, U^\dagger V^\dagger
\end{eqnarray}
where
\begin{eqnarray}
\label{j0}
U&=&\frac{1}{\sqrt{2}}\left(\begin{array}{cc}e^{-i M n_\mu x_\mu-i\alpha} 
& e^{i M n_\mu x_\mu+i\alpha}\\-e^{-i M n_\mu x_\mu-i\alpha} & 
e^{i M n_\mu x_\mu+i\alpha}\end{array}\right)\nonumber\\
V&=&\left[A_k^2(-D^2)+B^2(-D^2)\right]^{-1/2}
\left(\begin{array}{cc}A_k(-D^2) & -B(-D^2)\\
B(-D^2) & A_k(-D^2)\end{array}\right)
\end{eqnarray}
with
\begin{eqnarray}
\label{-a28}
& &A_k(-D^2)\equiv2\omega_k(M^2)+\sqrt{4\omega_k^2(M^2)+B^2(-D^2)}\;,\;\;
B(-D^2)\equiv\Omega(-D^{*2})-\Omega(-D^2)\;,\nonumber\\
& &D_\mu\equiv\partial_\mu+i M n_\mu\;,\;\;D_\mu^*\equiv\partial_\mu-i M
n_\mu\,,
\end{eqnarray}
we find
\begin{eqnarray}
\label{-a24}
W_k^{n,\alpha}[J]&\equiv &\frac{1}{2}\,\int d^dx\,
\Big\{K_1(J;x,n,\alpha)\,\Lambda_{k,1}(-D^2)^{-1}\,K_1(J;x,n,\alpha)\nonumber\\
& &+K_2(J;x,n,\alpha)\,\Lambda_{k,2}(-D^2)^{-1}\,K_2(J;x,n,\alpha)\Big\}
+{\cal O}(J^2J^{*2})\,.
\end{eqnarray}
Here the operators
\begin{eqnarray}
\label{-a26}
\Lambda_{k,1/2}(-D^2)\equiv\omega_k(-D^2)+\omega_k(-D^{*2})-4
\omega_k(M^2)\mp\sqrt{4\omega_k^2(M^2)+B^2(-D^2)}
\end{eqnarray}
represent the inverse propagators for the {\it real} fields $K_1$ and $K_2$ 
which are 
defined as
\begin{eqnarray}
\label{-a25}
K_1(J;x,n,\alpha)&\equiv &\left[A_k^2(-D^2)+B^2(-D^2)\right]^{-1/2}
\bigg\{ A_k(-D^2)\,{\rm Re}\left(J(x)e^{-iM n_\mu x_\mu-i
\alpha}\right)\nonumber\\
& &+i B(-D^2)\,{\rm Im}\left(J(x)e^{-iM n_\mu x_\mu-i\alpha}\right)\bigg\}
\nonumber\\
K_2(J;x,n,\alpha)&\equiv &\left[A_k^2(-D^2)+B^2(-D^2)\right]^{-1/2}
\bigg\{ A_k(-D^2)\,{\rm Im}\left(J(x)e^{-iM n_\mu x_\mu-i
\alpha}\right)\nonumber\\
& &+i B(-D^2)\,{\rm Re}\left(J(x)e^{-iM n_\mu x_\mu-i\alpha}\right)\bigg\}\,.
\end{eqnarray}
Since we dropped the integration over the vacuum manifold, we can now perform 
the Legendre-transformation directly on (\ref{-a24}) so that the analogue of
the effective average action takes the form
\begin{eqnarray}
\label{-a27}
\Gamma_k^{n,\alpha}[\phi]&=&\frac{1}{2}\,\int d^dx
\,\Big\{\Phi_1(\phi;x,n,\alpha)\,
\left[\Lambda_{k,1}(-D^2)-2k^2\right]\,\Phi_1(\phi;x,n,\alpha)\nonumber\\
& &+\Phi_2(\phi;x,n,\alpha)\,
\left[\Lambda_{k,2}(-D^2)-2k^2\right]\,\Phi_2(\phi;x,n,\alpha)\Big\}
+{\cal O}(\phi^2\phi^{*2})\,.
\end{eqnarray} 
The relations between the {\it real} average fields $\Phi_1$ and $\Phi_2$ and 
the complex average field $\phi\equiv\delta W_k^{n,\alpha}/\delta J$ can be 
read off from 
eq. (\ref{-a25}) if one replaces $J$ with $\phi$ and $K_i$ with $\Phi_i$. 

Obviously the effective kinetic terms for the fields $\Phi_1$, $\Phi_2$ are
given by $\sigma_{k,1/2}(-D^2)\equiv\Lambda_{k,1/2}(-D^2)-2k^2$. After going
over to momentum space it is easy to see that the kinetic terms yield 
nonnegative expressions for $k=0$. In fact, $\sigma_{0,2}((p+Mn)^2)=
\Lambda_{0,2}((p+Mn)^2)>0$ for all $p$. Furthermore, $\sigma_{0,1}((p+Mn)^2)=
\Lambda_{0,1}((p+Mn)^2)>0$ for all $p\neq 0$, while $\sigma_{0,1}(M^2)=
\Lambda_{0,1}(M^2)=0$. Thus, for the vacuum consisting of a single plane
wave, all modes of the theory at $k=0$ are found to be stable.

For $k=0$, $\Gamma_k^{n,\alpha}$ coincides with the action $S_{\rm fluct}$
discussed in section \ref{2}. In momentum space we have
\begin{eqnarray}
\label{b5}
{\cal K}_j(p^2,\theta)=\Lambda_{k=0,j}\left(\left(p_\mu+M n_\mu\right)^2
\right)
\end{eqnarray}
with $\cos\theta=p_\mu n_\mu/|p|$.
\section{Effective kinetic term in three and four dimensions}
\label{C}
$d=3$:
\begin{eqnarray}
\label{-a21}
\lefteqn{\tilde{\Sigma}_{k\rightarrow 0}(|p|=Mq)}\nonumber\\
&=&M^2\left\{\frac{4\left(q^{10}-4q^8+7q^6-6q^4+2q^2\right)^{1/2}-\arctan\left[
\frac{q\left(q^6-4q^5+q^4+8q^3-4q^2-8q+6\right)}
{\left(q^{10}-4q^8+7q^6-6q^4+2q^2\right)^{1/2}}\right]}
{4\left(q^4-2q^2+2\right)\left(q^{10}-4q^8+7q^6-6q^4+2q^2\right)^{1/2}}\right.
\nonumber\\
& &+\left.\frac{4\left(q^{10}-4q^8+7q^6-6q^4+2q^2\right)^{1/2}+\arctan\left[
\frac{q\left(q^6+4q^5+q^4-8q^3-4q^2+8q+6\right)}
{\left(q^{10}-4q^8+7q^6-6q^4+2q^2\right)^{1/2}}\right]}
{4\left(q^4-2q^2+2\right)\left(q^{10}-4q^8+7q^6-6q^4+2q^2\right)^{1/2}}
\right\}^{-1}
\end{eqnarray}
$d=4$:
\begin{eqnarray}
\label{-a22}
\lefteqn{\tilde{\Sigma}_{k\rightarrow 0}(|p|=Mq)=
8M^2q^2\left\{\frac{16q^6-32q^4+32q^2-2}{\left(q^4-2q^2+2\right)^2}
+2(q-1)^{-1}\left(q^4-2q^2+2\right)^{-5/2}\right.}\nonumber\\
& &\times\left(\frac{\left(q^7-7q^6+13q^5
+5q^4-27q^3+5q^2+29q-19\right)^2}{(q+1)^2
\left(q^{12}-22q^{10}+163q^8-436q^6+731q^4-670q^2+361\right)}\right)^{1/2}
\nonumber\\
& &\times\left\{(q-1)\left(q^4-2q^2+2\right)^{1/2}\left[
\frac{1}{2}\frac{(q+1)(q^6-11q^4+21q^2-19)}{q^7-7q^6+13q^5+5q^4
-27q^3+5q^2+29q-19}\right.\right.\nonumber\\
& &\left.+\frac{1}{2}
\left(\frac{(q+1)^2\left(q^{12}-22q^{10}+163q^8-436q^6+731q^4-670q^2+361
\right)}{\left(q^7-7q^6+13q^5+5q^4-27q^3+5q^2+29q-19\right)^2}\right)^{1/2}
\right]^{1/2}\nonumber\\
& &+{\rm sign}\left(\frac{q\left(q^4-2q^2+2\right)^{1/2}}
{q^7-7q^6+13q^5+5q^4-27q^3+5q^2+29q-19}\right)\nonumber\\
& &\times\left(q^5+3q^4-2q^3-6q^2+2q+6
\right)\nonumber\\
& &\times\left[-\frac{1}{2}\frac{(q+1)(q^6-11q^4+21q^2-19)}{q^7-7q^6
+13q^5+5q^4-27q^3+5q^2+29q-19}\right.\nonumber\\
& &\left.\left.\left.+\frac{1}{2}
\left(\frac{(q+1)^2\left(q^{12}-22q^{10}+163q^8-436q^6+731q^4-670q^2+361
\right)}{\left(q^7-7q^6+13q^5+5q^4-27q^3+5q^2+29q-19\right)^2}\right)^{1/2}
\right]^{1/2}\right\}\right\}^{-1}
\end{eqnarray}
\section{The four-point function}
\label{E}
In this part of the appendix we calculate the coefficient $u^{(4)}_k(p)$ 
which represents the fourth-order contribution to the expansion (\ref{e7}) of
the effective potential $U_k$, valid for $A<A_{\rm triv}$. 

We start from the assumptions that $k<k_{\rm cr}$ and that the sources are 
plane waves $J=\varepsilon\,\exp(ip_\mu x_\mu+i\beta)$ satisfying $|p|\neq M$, 
$\varepsilon<\varepsilon_k(p^2)$. This implies that the global minimum is 
degenerate and corresponds to the field $\chi_{\rm min}$ given by the expansion
(\ref{a42}) of appendix \ref{A}. Then it follows from section \ref{6} that, 
in the classical approximation, the generating functional $W_k$ is obtained 
from
\begin{eqnarray}
\label{h6}
\exp\left\{W_k[J]\right\}=N_k\,\int\limits_0^{2\pi}d\alpha\int 
d\mu(n)\,\exp\left\{-S_k^J\left[\chi_{\rm min};J\right]\right\}\,.
\end{eqnarray}
Let us consider the l.h.s. of this equation first. Because of 
{\sf U(1)}-invariance
and analyticity in $J$ and $J^*$, which we assume for sufficiently small $|J|$,
$W_k$ has an expansion of the form
\begin{eqnarray}
\label{h0}
W_k\left[J=\varepsilon\,\exp(ip_\mu x_\mu+i\beta)\right]
={\cal V}\left[-C_k+\left(\Sigma_k(p^2)+k^2\right)^{-1}\,\varepsilon^2
+G_k^{(4)}(p^2)\,\varepsilon^4+{\cal O}(\varepsilon^6)\right]
\end{eqnarray}
where the first two coefficients were already determined in section \ref{6}.
We will deal with the four-point function $G_k^{(4)}(p^2)$ for the
rest of this section. Note that it depends here only on one single momentum, 
$p$, which is due to the fact that we have inserted plane-wave sources.

Next we also expand the exponent on the r.h.s. of eq. (\ref{h6}) up to the
fourth order in $\varepsilon$. By a lengthy calculation (which we omit here 
for the sake of simplicity, except for $n=4$, see below) one finds that the 
corresponding coefficients consist of terms proportional to ${\cal V}$ and 
terms containing $\delta$-functions of momenta. Since the latter terms do not
contribute we obtain
\begin{eqnarray}
\label{h8}
\left.\frac{\partial^m}{\partial\varepsilon^m} S_k^J\left[\chi_{\rm min};
J=\varepsilon\,\exp(ip_\mu x_\mu+i\beta)\right]
\right|_{\varepsilon=0}
\equiv {\cal V}\,s_k^{(m)}(p,n;\alpha,\beta)
\end{eqnarray}
at least for $m=0,\ldots,4$.
Thus eq. (\ref{h6}) takes the form
\begin{eqnarray}
\label{h9}
\lefteqn{\exp\left\{{\cal V}\left[-C_k+\left(\Sigma_k(p^2)+k^2\right)\,
\varepsilon^2
+G_k^{(4)}(p^2)\,\varepsilon^4+{\cal O}(\varepsilon^6)\right]\right\}}
\nonumber\\
&=&N_k\,\int\limits_0^{2\pi}d\alpha\int d\mu(n)\exp\left\{-{\cal V}\sum
\limits_{m=0}^4\frac{1}{m!}\,s_k^{(m)}(p,n;\alpha,\beta)\,\varepsilon^m
+{\cal O}(\varepsilon^5)\right\}
\end{eqnarray}
Expanding both sides of (\ref{h9}) with respect to $\varepsilon$ and 
${\cal V}$ and comparing the coefficients of ${\cal V}\varepsilon^4$ then 
leads to
\begin{eqnarray}
\label{h7}
G_k^{(4)}(p^2)=-\frac{N_k}{4!\,{\cal V}}\,\int\limits_0^{2\pi}d\alpha\int 
d\mu(n)\,\left.\frac{\partial^4}{\partial\varepsilon^4}
S_k^J\left[\chi_{\rm min};J=\varepsilon\,\exp(ip_\mu x_\mu+i\beta)\right]
\right|_{\varepsilon=0}\,.
\end{eqnarray}
In order to deduce an explicit expression for the integrand of (\ref{h7})
we insert the expansion for the nontrivial, degenerate minimum (\ref{a42})
into $S_k^J$ and find 
\begin{eqnarray}
\label{h10}
\lefteqn{\left.\left.\frac{\partial^4}{\partial\varepsilon^4}
 S_k^J\left[\chi_{\rm min};J=\varepsilon\,\exp(ip_\mu x_\mu+i\beta)\right]
\right|_{\varepsilon=0}
=12\lambda\int d^dx\right\{\left|\varphi^{(1)}\right|^4}\nonumber\\
& &\left.+2\,\left|\varphi^{(1)}\right|^2
\,\left(\varphi^{(2)*},\varphi^{(2)}\right)\left(\begin{array}{c}\varphi^{(0)}
\\\varphi^{(0)*}\end{array}\right)
+\left(\varphi^{(2)*},\varphi^{(2)}\right)\left(\begin{array}{c}
(\varphi^{(1)})^2\,\varphi^{(0)*}
\\(\varphi^{(1)*})^2\varphi^{(0)}\end{array}\right)\right\}
\end{eqnarray}  
where $\varphi^{(0)}$ and $\varphi^{(1)}$ are determined by (\ref{a50}).
$\varphi^{(2)}$ can be derived from
\begin{eqnarray}
\label{h11}
\left.\frac{\partial^2}{\partial\varepsilon^2}
\frac{\delta S_k[\chi_{\rm min}]}{\delta\chi^*}\right|_{\varepsilon=0}=0
\end{eqnarray}
which represents the quadratic part of the e.o.m. (\ref{a30}) and leads to
\begin{eqnarray}
\label{h12}
\left(\begin{array}{c}\varphi^{(2)}\\\varphi^{(2)*}\end{array}\right)
=-\lambda\,\hat{S}_k^{(2)}\left[\varphi^{(0)}\right]^{-1}
\left[2\left|\,\varphi^{(1)}\right|^2\,\left(\begin{array}{c}\varphi^{(0)}
\\\varphi^{(0)*}\end{array}\right)
+\left(\begin{array}{c}(\varphi^{(1)})^2\,\varphi^{(0)*}
\\(\varphi^{(1)*})^2\varphi^{(0)}\end{array}\right)\right]\,.
\end{eqnarray}
After inserting (\ref{h12}) into (\ref{h10}) the resulting expression just 
depends on the fields $\varphi^{(0)}$ and $\varphi^{(1)}$. Using the explicit
expressions for these fields, (\ref{h7}) eventually yields
\begin{eqnarray}
\label{h13}
\lefteqn{\widetilde{G}_{k=M\kappa}^{(4)}(|p|=Mq)}\nonumber\\
&=&\frac{\lambda}{2M^8}\frac{\Omega_{d-2}}
{\Omega_{d-1}}\int\limits_0^{(1+
\delta_{d,2})\pi}d\theta\,\sin^{d-2}\theta\Bigg\{
\left[\left(\frac{1}{2}-\kappa^2+\frac{1}{2}\left(q^2-1\right)^2\right)\,
h(q,\theta,\kappa)
-\left(\kappa^2-\frac{1}{2}\right)^2\right]^{-4}\nonumber\\
& &\times\Bigg\{3\,h(q,\theta,\kappa)^4
+8\left(\kappa^2-\frac{1}{2}\right)\,h(q,\theta,\kappa)^3
+8\left(\kappa^2-\frac{1}{2}\right)^2\,h(q,\theta,\kappa)^2\nonumber\\
& &+8\left(\kappa^2-\frac{1}{2}\right)^3\,h(q,\theta,\kappa)
+3\left(\kappa^2-\frac{1}{2}\right)^4\nonumber\\
& &-\left(\kappa^2-\frac{1}{2}\right)
\bigg[-\left(\kappa^2-\frac{1}{2}\right)^2+\left(\frac{1}{2}-\kappa^2+8q^2
\cos^2\theta
-16q^3\cos\theta+8q^4\right)\nonumber\\
& &\times\left(\frac{65}{2}-\kappa^2-96q\cos\theta+\left(32+72
\cos^2\theta
\right)q^2-48q^3\cos\theta+8q^4\right)\bigg]^{-1}\nonumber\\
& &\times
\Bigg[2\left(\frac{65}{2}-\kappa^2-96q\cos\theta+\left(32+72\cos^2\theta
\right)q^2-48q^3\cos\theta+8q^4\right)\,h(q,\theta,\kappa)^4\nonumber\\
& &+8\left(\kappa^2-\frac{1}{2}\right)\,\left(32-96q\cos\theta+\left(32+72
\cos^2\theta\right)q^2-48q^3\cos\theta+8q^4\right)\,h(q,\theta,\kappa)^3
\nonumber\\
& &+8\left(\kappa^2-\frac{1}{2}\right)^2\,\Bigg(\frac{127}{4}+\frac{\kappa^2}
{2}-96q\cos\theta+\left(32+80\cos^2\theta\right)q^2\nonumber\\
& &-64q^3\cos\theta+16q^4\Bigg)\,h(q,\theta,\kappa)^2\nonumber\\
& &+8\left(\kappa^2-\frac{1}{2}\right)^3\,\left(8q^2\cos^2\theta-16q^3
\cos\theta+8q^4\right)\,h(q,\theta,\kappa)\nonumber\\
& &+2\left(\kappa^2-\frac{1}{2}\right)^4\left(\frac{1}{2}-\kappa^2+8q^2
\cos^2\theta-16q^3\cos\theta+8q^4\right)\Bigg]\Bigg\}\Bigg\}
\end{eqnarray}
where $G_k^{(4)}(p^2)\equiv \widetilde{G}_k^{(4)}(|p|)$ and
\begin{eqnarray}
\label{h14}
h(q,\theta,\kappa)\equiv
5-\kappa^2-12q\cos\theta+\left(3+8\cos^2\theta\right)q^2
-4q^3\cos\theta+\frac{q^4}{2}\,.
\end{eqnarray}
Since the integral over $\theta$ can be evaluated numerically for fixed values
of $p$ and $k$ an explicit expression for $G_k^{(4)}(p^2)$ is at our 
disposal and therefore also for the expansion of $W_k$ given by eq. (\ref{h0}).

Introducing $\phi=A\,\exp(ip_\mu x_\mu+i\beta)$ and applying the definition
\begin{eqnarray}
\label{h3}
{\cal V}\,U_k(A;p^2)&=&\Gamma_k\left[\phi=A\,\exp(ip_\mu x_\mu+i\beta)\right]
\nonumber\\
&=&{\cal V} \left[2A\,\varepsilon(A)-k^2\,A^2\right]-W_k\left[J=\varepsilon(A)
\,\exp(ip_\mu x_\mu+i\beta)\right]
\end{eqnarray}
we may now determine the corresponding approximate effective potential $U_k$
as described in section \ref{8}.
The relation $\varepsilon(A)$ appearing in (\ref{h3}) is obtained by
inverting
\begin{eqnarray}
\label{h1}
A&=&\frac{1}{2{\cal V}}\frac{\partial}{\partial\varepsilon}
W_k\left[J=\varepsilon\,\exp(ip_\mu x_\mu+i\beta)\right]\nonumber\\
&=&\left[\left(\Sigma_k(p^2)+k^2\right)^{-1}\,\varepsilon+2\,G_k^{(4)}(p^2)\,
\varepsilon^3+{\cal O}(\varepsilon^5)\right]
\end{eqnarray}
which yields
\begin{eqnarray}
\label{h2}
\varepsilon(A)=\left(\Sigma_k(p^2)+k^2\right)\,A-2\left(\Sigma_k(p^2)
+k^2\right)^{4}\,G_k^{(4)}(p^2)\,A^3+{\cal O}(A^5)\,.
\end{eqnarray}
Inserting this expression into (\ref{h3}) finally leads to
\begin{eqnarray}
\label{h4}
U_k(A;p^2)=C_k+\Sigma_k(p^2)\,A^2-\left(\Sigma_k(p^2)+k^2\right)^4 G_k^{(4)}
(p^2)\,A^4+{\cal O}(A^6)\,.
\end{eqnarray}
Thus we can identify the coefficient $u_k^{(4)}(p)$ as
\begin{eqnarray}
\label{h5}
u_k^{(4)}(p)=-\left(\Sigma_k(p^2)+k^2\right)^4 G_k^{(4)}(p^2)\,.
\end{eqnarray}
\end{appendix}


\begin{thebibliography}{99999999}
\bibitem{sav}G. K. Savvidy, {\it Infrared instability of the vacuum state of
gauge theories and asymptotic freedom}, Phys. Lett. B 71 (1977), 133-134
\bibitem{haw}See, for instance, S. W. Hawking in {\it Relativity, Groups and
Topology II}, Proceedings of the Les Houches Summer School 1983, B. S. DeWitt,
R. Stora, Eds., North Holland, Amsterdam, 1984
\bibitem{liou}E. D'Hoker, R. Jackiw, {\it Classical and quantal Liouville 
field theory}, Phys. Rev. D 26 (1982), 3517-3542;
{\it Space-Translation Breaking and Compactification in the Liouville Theory},
Phys. Rev. Lett. 50 (1983), 1719-1722
\bibitem{liou2}A. H. Chamseddine, M. Reuter, {\it Induced two-dimensional 
quantum gravity and SL(2,R) Kac-Moody current algebra}, Nucl. Phys. B 317 
(1989), 757-771;\\
M. Reuter, C. Wetterich, {\it Quantum Liouville field theory
as solution of a flow equation}, Nucl. Phys. B 506 (1997), 483-520, 
hep-th/9605039
\bibitem{avact}C. Wetterich, {\it Exact evolution equation for the effective 
potential}, Phys. Lett. B 301 (1993), 90-94;\\
M. Reuter, C. Wetterich, {\it Effective average action for gauge theories and 
exact evolution equations}, Nucl.Phys. B 417 (1994), {\it } 181-214;\\
M. Reuter, C. Wetterich, {\it Exact evolution equation for scalar 
electrodynamics}, Nucl. Phys. B 427 (1994), 291-324;\\
for an introduction see: M. Reuter, {\it Effective Average Actions and 
Nonperturbative Evolution Equations}, hep-th/9602012, to appear in the 
Proceedings of the {\it 5th Hellenic School and Workshops on Elementary 
Particle Physics}, Corfu, Greece, 1995
\bibitem{RW90}A. Ringwald, C. Wetterich, {\it Average Action for the N 
Component $\phi^4$ Theory}, Nucl. Phys. B 334 (1990), 506-526
\bibitem{ABP99}J. Alexandre, V. Branchina, J. Polonyi, {\it Instability
Induced Renormalization}, Phys. Lett. B 445 (1999), 351-356, cond-mat/9803007 
\bibitem{TW92}N. Tetradis, C. Wetterich, {\it Scale dependence of the average
potential around the ma\-xi\-mum in $\phi^4$-theories}, Nucl. Phys. B 383 
(1992), 197-217;\\
J. Berges, N. Tetradis, C. Wetterich, {\it Nonperturbative renormalization flow
in quantum field theory and statistical physics}, hep-ph/0005122
\bibitem{BMP}J. Polonyi, {\it The antiferromagnetic vacuum}, hep-lat/9610030;\\
V. Branchina, H. Mohrbach, J. Polonyi, {\it The 
antiferromagnetic $\phi^4$ Model, I. The Mean-field solution}, Phys. Rev. D 60
(1999), 045006, hep-th/9612110;\\
V. Branchina, H. Mohrbach, J. Polonyi, {\it The 
antiferromagnetic $\phi^4$ Model, II. The one-loop renormalization}, Phys. Rev.
D 60 (1999), 045007, hep-th/9612111 
\bibitem{wipfetal}L. O'Raifeartaigh, A. Wipf, H. Yoneyama, {\it The constraint
effective potential}, Nucl. Phys. B 271 (1986), 653-680
\bibitem{CW}S. Coleman, E. Weinberg, {\it Radiative corrections as the origin
of spontaneous symmetry breaking}, Phys. Rev. D 7 (1973), 1888-1910
\bibitem{aagrav}M. Reuter, {\it Nonperturbative Evolution Equation for
Quantum Gravity}, Phys. Rev. D 57 (1998), 971-985, hep-th/9605030
\bibitem{nonloc}C. Wetterich, {\it Effective Nonlocal Euclidean Gravity}, Gen.
Rel. Grav. 30 (1998), 159-172, gr-qc/9704052
\bibitem{WRW}C. Wetterich, {\it Spontaneous compactification in higher 
dimensional gravity}, Phys. Lett. B 113 (1982), 377-381;\\
M. Reuter, C. Wetterich, {\it Classical stability for spontaneous 
compactification in higher derivative gravity}, Nucl. Phys. B 289 (1987),
757-786
\end{thebibliography}
\end{document}